# Factors that Affect Software Systems Development Project Outcomes: A Survey of Research

Laurie McLeod and Stephen G. MacDonell

*SERL, Auckland University of Technology*
*Private Bag 92006, Auckland 1142*
*New Zealand*
*{laumcl88, stephen.macdonell}@aut.ac.nz*

**Abstract**

*Determining the factors that have an influence on software systems development and deployment project outcomes has been the focus of extensive and ongoing research for more than 30 years. We provide here a survey of the research literature that has addressed this topic in the period 1996-2006, with a particular focus on empirical analyses. On the basis of this survey we present a new classification framework that represents an abstracted and synthesized view of the types of factors that have been asserted as influencing project outcomes.*

**Keywords**: *Development processes; institutional context; people and action; project content; project outcomes.*

## 1. INTRODUCTION

Such has been the scale of investment of time, effort and money in software systems that achieving success in their development and deployment has consumed many researchers (and of course, practitioners) in the last three decades. As far back as 1975 researchers were attempting to distil the determinants of project success and failure so that a successful outcome was more likely [Lucas, 1975] – with mixed results. In the early days we were in effect attempting – or perhaps hoping – to control something we did not understand. Since the mid-1980s we have acquired a far more detailed understanding of why and how systems are developed; ensuring success (by some definition) remains a significant challenge, however. There are many, many factors that are said to have an effect on software systems development project outcomes – some, researchers have speculated upon, others have been demonstrated empirically. In this paper we focus on the latter in an effort to identify and classify the factors that have been shown to have an impact on project outcomes.

We first briefly consider previous classifications of the relevant literature as a benchmark for our own effort. A classificatory framework that emerged from our synthesized analysis of empirical studies of software systems development project outcomes is then presented, followed by a description of the approach used to conduct the empirical review. In Section 3, we consider what researchers actually mean by 'project outcomes' – use of such a descriptor is widespread but in many instances it is left undefined, the often unjustified assumption being that its meaning is somehow 'obvious'. Sections 4 through 7 of the paper then report on the (non-exclusive) classes of potentially influential factors that have been described in empirical studies published in academic journals between 1996 and 2006. Each self-contained section can be read separately, or as part of the whole. The paper closes with a discussion of the main themes emerging from the extensive literature review conducted in this study, followed by a short conclusion.

## 2. CLASSIFYING THE EMPIRICAL RESEARCH LITERATURE ON FACTORS INFLUENCING PROJECT OUTCOMES

Software systems development and deployment projects and the influences shaping their outcomes have been the subjects of a sustained research effort over the last thirty-five years. Making sense of the huge volume of empirical findings on this topic is a daunting task.

A number of authors have attempted to organize the research in this area (Table 1). A classification of reasons for systems failure produced by Lyytinen & Hirschheim [1987] was an early attempt to provide a systemic framework based on an extensive empirical survey. Since 1987, other authors have produced various classificatory frameworks or models in their attempts to develop an understanding of systems success or failure. (We do not define success and failure here – this is addressed in Section 3.) For example, Poulymenakou & Holmes [1996] utilize a contingency model to examine the effects of various macro and micro contextual factors on systems project processes and outcomes. Butler & Fitzgerald [2001] also focus on the effect of institutional contexts on the systems development process and systems success. They delineate a range of project-, process-, and user-related factors, the content and conduct of which influence development outcomes through user participation and management of change. Synthesizing literature on critical success factors in enterprise system implementations and software and IT project risks, Scott & Vessey [2002] develop a model of enterprise resource planning (ERP) project risk factors based on four-levels of forces, both external and internal to the organization, influencing changes arising from organizational adoption of enterprise systems.

While of value, the classificatory schemes summarized in Table 1 are limited in one or more ways. For example, the studies on which the Lyytinen & Hirschheim [1987] framework is based date from 1985 or earlier and may not reflect more recent changes in the nature of software systems development. Subsequent classificatory schemes are limited in their level of detail or scope, either focusing on specific aspects of systems development [Butler & Fitzgerald, 2001], a subset of influential factors [Poulymenakou & Holmes, 1996], or specific types of systems projects [Scott & Vessey, 2002].

This paper redresses these limitations by offering a more current survey of empirical work on systems development, while still recognizing the desirability of a systematic conceptualization of the field emphasized by Lyytinen & Hirschheim [1987]. It synthesizes the findings of empirical studies that address a wide range of project outcomes, including both systems success and failure, as well as project performance or abandonment. The classificatory framework developed in this paper builds on the earlier examples described above through an explicit focus on features of the project itself, a consideration of the process of software systems development, together with the actions of human agents, and the inclusion of a range of contextual influences, spanning individual, development, organizational and environmental levels. This focus on project content, process, action and context reflects the underlying conceptualization of the new classificatory framework outlined below.

**Table 1:** Prior schemes classifying influences on software systems development and deployment project outcomes

| Study | Scope | Categories (as defined in original study) |
|---|---|---|
| Lyytinen & Hirschheim [1987] | Classificatory framework of reasons for system failure | 1. Features of the system – technical and operational<br>2. Features of the system environment – lack of fit of the system with users, rest of organization, or operating environment<br>3. Features of the systems development process – methods, decision-making, assumptions, nature of work, implementation, contingency factors<br>4. Features of systems development environment – cognitive and skill limitations of analysts or users |
| Poulymenakou & Holmes [1996] | Contingency framework of variables affecting system failure | 1. Macro (organizational) contingent variables – culture, planning, accountability, irrationality, evaluation<br>2. Micro (project) contingent variables – power and politics, user resistance to change, development methods |
| Butler & Fitzgerald [2001] | Model for user participation and management of change in systems development | 1. Institutional context – organizational policy<br>2. Project-related factors – project initiator, top management commitment, type of system, project complexity, task-structure, time for development, available financial resources, resultant change<br>3. Process-related factors – user-developer relationship, influence and power, communication<br>4. User-related factors – user perceptions, commitment, willingness, ability, characteristics and attitudes |
| Scott & Vessey [2002] | Model of risk factors in ERP system implementations | 1. External business environment<br>2. Organizational context – culture, structure, strategy, IT infrastructure, business processes<br>3. Systems context – data, technology, project governance<br>4. Project – project focus and scope, project management, change management |

## 2.1 A New Classificatory Framework

The classificatory framework presented in this paper was developed as a way of synthesizing and organizing contemporary empirical knowledge of the various influences that affect systems development outcomes, and in order to provide a basis for a more systematic understanding of the factors that reportedly produce various system outcomes [Lyytinen & Hirschheim, 1987]. The framework is the outcome of a conceptual understanding of software systems development, informed by the results of a subsequent review of empirical literature. Although the final form of the framework emerged after the synthesis of the empirical literature, it is presented first in this paper in order to provide conceptual assistance to the reader in making sense of the breadth of information presented on software systems development [Gallivan & Keil, 2003].

Walsham's [1993] theoretical treatment of the content, process and context of system-related organizational change was used as a conceptual basis for organizing the findings of the prior empirical research. Walsham [1993] suggests that a consideration of all three of these components in systems design, development and implementation overcomes a prior over-emphasis on the content of systems-related change at the expense of the process of change and its relationship with organizational and wider contexts. The content, process and context concepts have been used elsewhere in the software systems literature, suggesting a degree of recognition and acceptance [Kautz, 2004; Kautz & Nielsen, 2004; Serafeimidis & Smithson, 1999; Stockdale & Standing, 2006]. In addition, these concepts are sufficiently generic to accommodate the wide variety of proposed factors influencing systems project outcomes, while providing a parsimonious framework for reviewing and classifying them [Stockdale & Standing, 2006].

To the content, process and context of system-related organizational change, we add a fourth component based on people and action. An explicit treatment of people and their actions is warranted given the increased recognition of software systems development as a social process [e.g. Flynn & Jazi, 1998; Kirsch & Beath, 1996; Robey et al., 2001]. Indeed, the social is an integral part of Walsham's [1993] understanding of context and process, and he emphasizes that "computer-based information systems are associated with organizational stability and change directly through the agency of human actors" [p. 253]. In this respect we are making explicit a dimension that was implicit in Walsham's original analytical framework.

The resultant framework is shown in Figure 1[1]. Influences on software systems development are divided into four major dimensions, each containing a number of factors highlighted in the subsequent review of empirical studies of systems development. The interaction of these four dimensions influences the project trajectory at particular points in time, including **project outcomes**. However, the evaluation of project outcomes is problematic given their multi-dimensional nature, the different evaluative criteria used by various stakeholders, and their often negotiated or contested nature.

**Project content** includes factors that are typically considered as properties of the software systems project itself, including its dimensions, scope and goals, the resources it attracts, and the hardware and software used in development. The project's properties, whether technological, strategic or resource-related, materially influence the development outcome or are mobilized and drawn upon by various individuals or groups in their development activities and interactions with each other.

**Development processes** include the various activities typically associated with software systems development and deployment, ranging from requirements determination and standard method use to the management of change resulting from implementation of a new software system. Grounded in a long history of received wisdom, these processes reflect the ongoing development of systems development practice. They constitute both opportunities and sites for action and interaction among the interested individuals and groups as they negotiate a particular system outcome.

**Institutional context** includes factors related to both the organization in which software systems development is located and the wider socio-economic environment in which the organization operates. A range of contextual properties or conditions, operating across local, national and international levels, and including the historical circumstances from which software systems are developed and used, may shape the course of development in a software systems project by constraining or facilitating certain courses of action.

Finally, as can be drawn from the content, context and process descriptions above, consideration needs to be given to the various **people**, both individuals and groups, who are involved or interested in the software systems development project. Their characteristics, **actions**, interactions and relationships shape the development trajectory and project outcomes in multiple ways, so an understanding of their roles and actions during system development is also necessary.

The classificatory framework developed here serves as an analytical device to facilitate understanding and order out of the vast amount of information available on factors influencing the development and deployment of software systems. The framework does not attempt to privilege any factor or set of factors over others. Project outcomes are highly situational in nature. While generic influences are common to a range of systems development contexts, they manifest themselves differently in specific situations [cf. Poulymenakou & Holmes, 1996].

Further, it is acknowledged that many of these factors are interrelated and that often project outcomes involve multiple factors that interact in complex ways. Indeed, the theoretical concepts of the content, process and context of change are typically treated as interlinked and in continuous interplay [Kautz, 2004; Kautz & Nielsen, 2004; Serafeimidis & Smithson, 1999; Stockdale & Standing, 2006]. This is represented in Figure 1 using overlapping circles to reflect the interaction between project content, development processes and people's actions during development, all situated within the broader context in which software systems development occurs. The distinction made between these dimensions is for analytical purposes. In practice, all four dimensions are interrelated and mutually interactive. For example, the development of technical requirements for a software systems project occurs as part of the requirements process (a development process). However, the adequacy of their definition relies on the technical expertise of any developers involved (people and action). The technical requirements themselves will depend on the nature of the project itself, the existing organizational IT infrastructure, and any other internal systems with which the new system must integrate (project content). A range of external conditions may potentially influence the technical requirements, such as legislative requirements or technical constraints imposed by external entities such as government agencies or business partners who require access to data from the new system (institutional context).

Table 2 compares the factors comprising the classificatory framework developed here with those from the prior classificatory schemes shown in Table 1. As can be seen, no one prior scheme provides the comprehensive and systematic coverage offered by the new framework. Further, in most cases, coverage of factors within the specific dimensions – project content, development processes, institutional context, and people and action – is incomplete. The new framework therefore provides the basis for a contemporary, inclusive and holistic analysis of influential factors in systems development across multiple levels of analysis and accommodating recent changes in development methods and practices.

---

[1] While the use of overlapping circles to represent various dimensions resembles Stockdale & Standing's [2006] model for evaluating information systems, the subject and detail of the two frameworks differ.

**Figure 1:** A framework for understanding influences on software systems development and deployment project outcomes

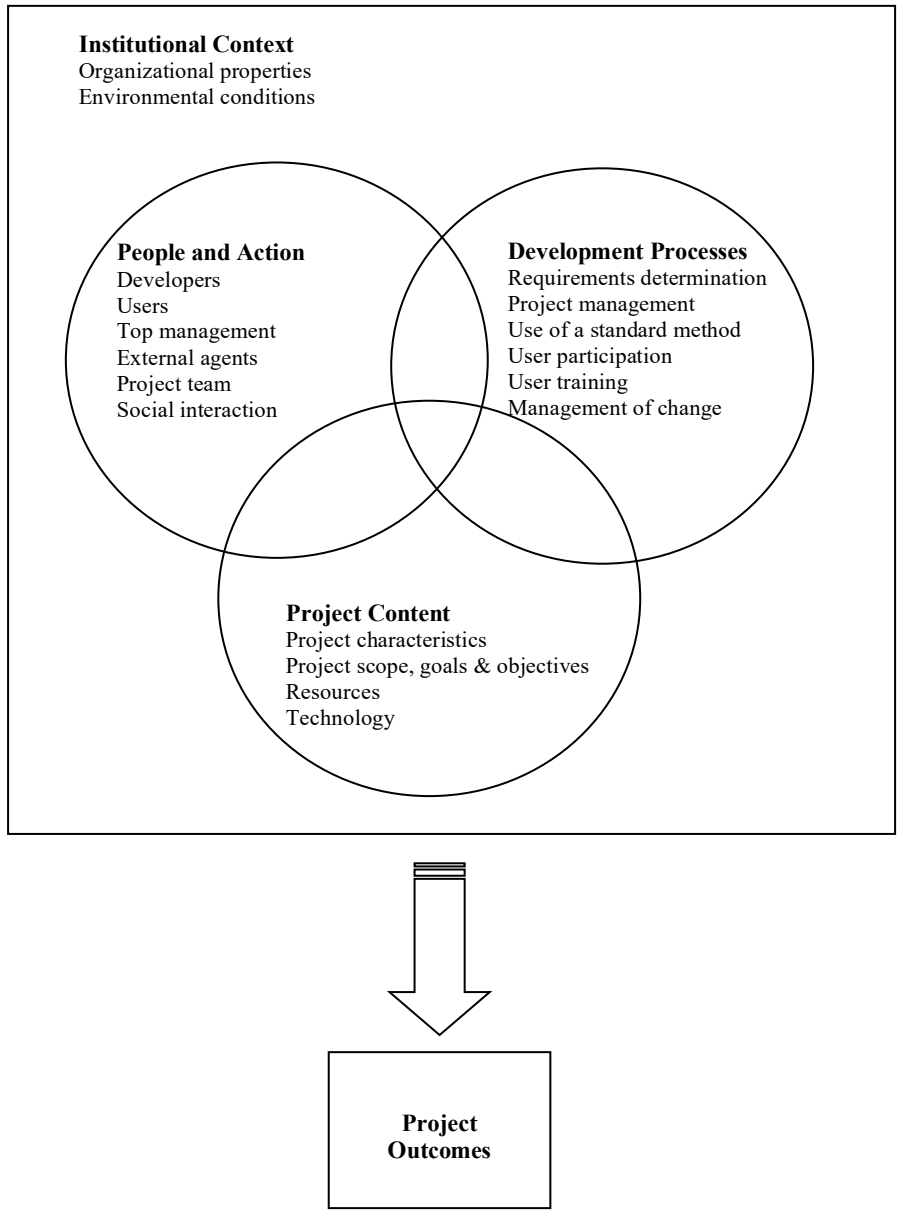

## 2.2 An Updated Review of Systems Development Research

As noted earlier, the classificatory framework presented here was used to synthesize an extensive survey of empirical literature on software systems development, addressing systems project outcomes and the wide variety of factors potentially influencing them. The literature review and synthesis also helped to refine and populate the initial conceptual classificatory framework.

Our focus is intentionally on *software systems*. By this, we mean software embedded in a context, normally organizational but sometimes individual. As a result we are interested primarily in studies that address system, rather that specifically software, development factors and outcomes. These software systems might be referred to as *information systems* or *management information systems* elsewhere; however, these labels can carry with them perhaps dated implications of transaction processing and reporting systems employed solely in a business context. The types of systems now prevalent – distributed systems, enterprise systems, systems utilizing web services and the like – may not be considered under such a categorization, but they are certainly of interest here. Hence we prefer the more inclusive *software systems* label. That said, studies that addressed the development of operating systems, real-time embedded systems and infrastructural systems are generally not covered here. This did not come about as the result of a deliberate strategy; rather, the search process did not lead to the retrieval of studies of this nature.

The scope of the review was studies of software systems development published between 1996 and 2006. This period was chosen in order to provide a basis for development of a contemporary classificatory framework. We primarily focused our search attention on archival research as published in journals. As a result we did not set out to address material published in conference proceedings, although some of this material did emerge from secondary searching. While this does exclude some of the research literature it also means that in general the bulk of the studies have been rigorously peer-reviewed and are considered to be of substantial archival value.

Each study included in the review was examined to establish that it met the following criteria:

- The study was published between 1996 and 2006.
- The study focused on software systems project outcomes or factors influencing software systems development (rather than system evaluation or systems in use).
- The study provided empirical data on systems project outcomes or factors influencing systems development.

An initial sample of studies was obtained using two EBSCO Information Services databases, *Computer Source* and *Business Source Premier*. *Computer Source* offers full text access to some 300 publications in computer science, information systems and software. *Business Source Premier* (over 2,300 full text journals) was included to provide coverage of journals in business and management information systems not covered by *Computer Source*. The two databases were searched between 1996 and 2006 using the terms 'project failure' or 'project success', in combination with the database-supplied subject terms 'computer software' or 'information technology'. These searches provided an initial total of 289 publications, many of which were eliminated from the study on the basis of the criteria outlined above. The remaining studies were reviewed and further candidate studies were identified from their reference lists using a 'snowballing' technique. In addition, a number of journals particularly relevant to the development of software systems but not covered in the initial database searches were subjected to targeted searches using the same search terms, resulting in the identification of a small number of additional studies. The aim was to be as inclusive as possible and the result was a comprehensive (although not exhaustive) survey of empirical research on factors influencing software systems development outcomes.

**Table 2:** Comparison of the new framework with prior classificatory schemes

| Factors Addressed | Lyytinen & Hirschheim [1987] | Poulymenakou & Holmes [1996] | Butler & Fitzgerald [2001] | Scott & Vessey [2002] |
|---|---|---|---|---|
| **People and Action** | | | | |
| Developers | ✓ | | | |
| Users | ✓ | ✓ | ✓ | |
| Top management | | | ✓ | ✓ |
| External agents | | | | ✓ |
| Project team | | | | ✓ |
| Social interaction | | ✓ | ✓ | ✓ |
| **Project Content** | | | | |
| Project characteristics | ✓ | | ✓ | ✓ |
| Project scope, goals & objectives | ✓ | | | ✓ |
| Resources | | | ✓ | ✓ |
| Technology | ✓ | | | ✓ |
| **Development Processes** | | | | |
| Requirements determination | | | | |
| Project management | | ✓ | | ✓ |
| Use of a standard method | ✓ | ✓ | | |
| User participation | | | ✓ | |
| User training | | | | ✓ |
| Management of change | ✓ | ✓ | ✓ | ✓ |
| **Institutional Context** | | | | |
| Organizational properties | | ✓ | ✓ | ✓ |
| Environmental conditions | | | | ✓ |

The empirical studies reviewed used a variety of data collection methods, including surveys, interview programs, Delphi studies and case studies. The most common approach was a factor-based study utilizing a large-scale survey and focused on one or a combination of people-, process- or project content-related factors. The case studies included in the literature review proved to be a particularly important source of contextual factors influencing software systems development. Factor-based empirical studies were reviewed first. These studies generated a range of author-reported influential factors, which were then assigned to one of the four dimensions of the classificatory framework and grouped into categories within each dimension based on their commonalities. As more studies were reviewed emergent categories were refined or combined to produce the categories used to populate each dimension in the classificatory framework shown in Figure 1. Where available, statistical significance was used to identify influential factors affecting systems development project outcomes in empirical studies based on quantitative data. Other quantitative studies used rank order to indicate the relative importance of individual factors. In empirical studies using qualitative data, reliance was placed on the authors' descriptions of influential factors, particularly the frequency with which a factor was mentioned, the magnitude of the reported effect, or authors' self-reported estimates of relative importance.

It is beyond the physical constraints of a journal publication to summarize details of the focus, method, findings and measurement criteria for all the empirical studies reviewed in this paper. However, thirty of the largest factor-based studies, each focusing on multiple (at least eight) influential factors across at least three dimensions, are tabulated in Appendix A. These are supplemented in the following discussion by further empirical studies that focused on systems project outcomes or that concentrated on a smaller number of relevant factors or a specific aspect of software systems development. Altogether, 177 empirical papers were reviewed for this study. Additional recent conceptual or theoretical papers on software systems development and project outcomes are also cited in order to provide clarification or explanation of factors highlighted in the empirical studies.

The results of the literature review are presented below based on the various components of the classificatory framework shown in Figure 1. The analysis first considers the problematic nature of defining software systems development project outcomes (Section 3). It then proceeds to review the (sometimes inter-related) influence of factors associated with people and action (Section 4), project content (Section 5), systems development processes (Section 6) and the institutional context (Section 7). At the end of each section Figure 1 is revisited to highlight key aspects of the dimension being considered.

## 3. PROJECT OUTCOMES

In the literature on software systems development, the outcome of a project is typically conceived of in terms of whether the project is successful or not. Identifying just what constitutes 'success' or 'failure', however, can be problematic. In general, there remains a lack of consensus on how to define success, lack of success, and failure. Such terms are perceived to be vague and difficult to measure [Butler & Fitzgerald, 1997, 2001; Lynch & Gregor, 2004; Wilson & Howcroft, 2002].

It is generally recognized that success and failure are multi-dimensional constructs, with interrelated technical, economic, behavioral, psychological and political dimensions [Bussen & Myers, 1997; DeLone & McLean, 2003; Doherty et al., 2003; Lynch & Gregor, 2004; Wixom & Watson, 2001]. In order to make them somehow more tangible, project success and failure have been defined (and measured) in terms of the software systems development process and/or its product. That is, success is a high quality development process outcome and/or a high quality product outcome [Barki et al., 2001; Karlsen et al., 2005; Markus & Mao, 2004; Nelson, 2005; Procaccino & Verner, 2006; Wixom & Watson, 2001]. DeLone and McLean [2003], for example, describe product success in terms of system quality, information quality, services quality, use (or intention to use), user satisfaction and net benefits. In terms of the development process, some authors have described the outcome in terms of whether or not the project is completed on time or within budget [Standish Group International, 1999, 2001; Wixom & Watson, 2001]. Other authors define project outcomes in terms of whether a project is smoothly completed, redefined or abandoned [Yetton et al., 2000]. While systems projects will ideally have successful process *and* product outcomes, Wallace & Keil [2004] suggest that projects emphasizing process outcome goals (such as budget and schedule) will be managed differently to those emphasizing product-related outcomes.

Markus & Mao [2004] distinguish development success from implementation success, the latter of which they view as a change management process and/or outcome. They note that current system contexts may necessitate extending success beyond the system to a wider solution that includes the system *and* complementary business or process interventions (i.e. to solution success). They suggest that there is not necessarily a relationship between system or solution development success and system or solution implementation success. In a similar manner, Crowston et al. [2006] examine the increasing interest in open source software development and present additional concepts of software systems success that they suggest are more appropriate for this particular development domain.

A number of researchers have approached success or failure in terms of the ability of a software system to meet the expectations of its stakeholders. In Lyytinen and Hirschheim's [1987] view, system failure and success form a continuum in which the likelihood of fulfilling an individual's expectations varies from very low to very high. Furthermore, groups or individuals may differ in their assessments of the extent to which a system is successful, judging it according to different criteria. Moreover, their opinions and evaluative assessments are fluid and may change over time, in response to political manoeuvring, persuasion, or changes in the organizational and technological context [Briggs et al., 2003; Bussen & Myers, 1997; DeLone & McLean, 2003; Jiang & Klein, 2000; Jiang et al., 1998a; Karlsen et al., 2005; Kim et al., 1999/2000; Lynch & Gregor, 2004; Nelson, 2005; Skok & Legge, 2002; Standing et al., 2006; Wilson & Howcroft, 2002].

For example, in a study of a project in the social services sector, Riley & Smith [1997] found that the many different stakeholders identified had different assessments of system success. The authors categorized the stakeholders into (1) project team members, who believed the system was innovative and valuable; (2) those outside the project team who questioned the basis of the project but thought that the system was worthwhile; and (3) user groups, some of whom accepted the system and made it work, others of whom rejected it. In another example, some members from a team of systems developers interviewed by Linberg [1999] suggested that even a project that was not completed could be successful, so long as some learning occurred that could be carried forward to future projects.

Karlsen et al. [2005] suggest that project evaluations should reflect some of the issues outlined above. They recommend that success criteria need to be defined from the outset, using input from the various stakeholders and incorporating a range of criteria, and that they may need to be modified to reflect changes that occur during the course of a project. They also suggest that multiple evaluations should be undertaken at different points in time, for different purposes (e.g. a project management assessment could be done during project execution and in the project delivery stage, whereas a user assessment should be done after users had had some experience using the system).

A number of authors suggest that success or failure should be thought of as a process rather than a single discrete outcome [Wilson & Howcroft, 2002]. From this perspective, the success or failure of a software system is constructed as the result of negotiated or contested subjective interpretations, and needs to be viewed against the historical context of systems development and the complex social and political interactions it involves [Mitev, 2000; Wilson & Howcroft, 2000, 2002]. Wilson & Howcroft [2002] further note that apparent definitional closure may not necessarily represent consensus. Dissenters may be forced to accept the situation or be denied a legitimate voice, particularly if they are considered to be of lesser standing in the project. The authors describe a nursing system which was perceived as a success by its sponsors, but as a failure by its users. Three years after its implementation, with continued user resistance, poor performance and financial pressures, the system was finally acknowledged as a failure by its sponsors.

The above discussion reveals that labeling a project outcome as a 'success' or 'failure' can be both difficult and problematic. However, despite their definitional or conceptual ambiguity, these terms are still frequently used (and measured, often via proxy indicators) in systems development and deployment research. Project outcomes vary along a continuum, may be interpreted differently from different perspectives, and are in many cases constructed through processes of sense-making and negotiation with or within an organization. Use of the terms 'success' or 'failure' in the following discussion reflects their use by the authors of the empirical studies reviewed. We deliberately do not attempt to interpret (or re-interpret) the terms.

Figure 2 summarizes key aspects of the nature and definition of project outcomes.

**Figure 2:** Aspects of project outcomes

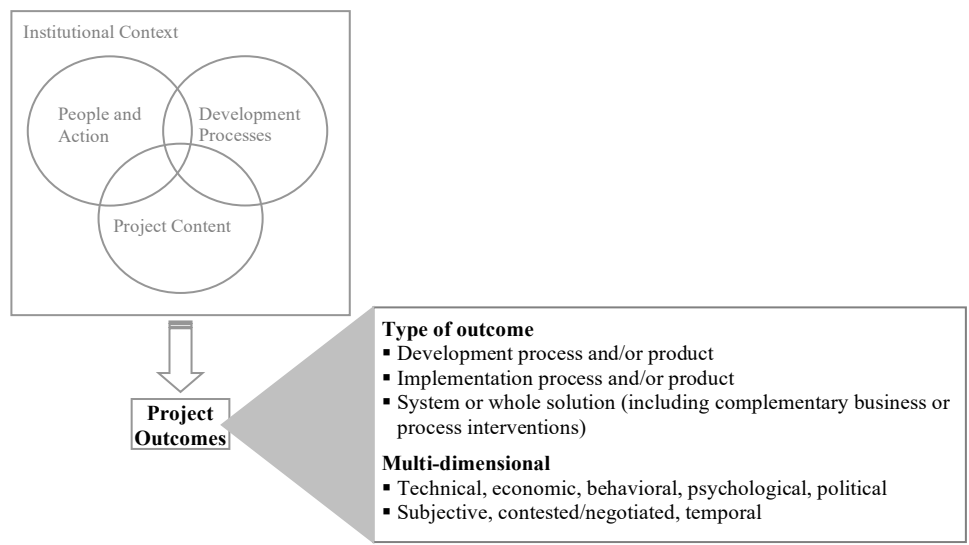

## 4. INFLUENTIAL FACTORS – PEOPLE AND ACTION

This section discusses influences related to the various individuals or groups with an interest in the system, their actions and interactions. Typical roles in software systems development include user, developer, development manager, business or user manager, project manager or leader, project team member, trainer, sponsor, customer, vendor, top management, and external consultant [Butler, 2003; Enquist & Makrygiannis, 1998; Heiskanen et al., 2000; Nandhakumar & Avison, 1999; Riley & Smith, 1997; Roberts et al., 2000; Robey & Newman, 1996; Robey et al., 2001]. In some cases, an individual may have multiple roles or their roles and perspectives may change over time [Pouloudi & Whitley, 1997; Robey et al., 2001]. It is also important to recognize that an individual's actions in pursuit of organizational objectives or in relation to a systems initiative will be influenced by competing commitments arising from social or professional groups he or she identifies with, organizational commitments institutionalized as organizational policy and practice, wider societal and cultural interests, or sectional interests arising as a result of the specific course of action [Butler, 2003]. For example, user managers can be conceived of as either users or managers. They can potentially mediate between their user group and higher level management [Marion & Marion, 1998], or they may be inclined to align themselves more with their user group.

To a large extent, the definition of such roles and the allocation of individuals to them are undertaken for analytical purposes and depend on the specific context and time frame under consideration [Pouloudi & Whitley, 1997]. Categorization into a particular group often reflects a researcher's preconceptions or bias rather than how individuals perceive themselves. The unreflective use of these groupings can be problematic for both research and practice [Butler & Fitzgerald, 1997; Iivari & Igbaria, 1997; van Offenbeek & Koopman, 1996]. The following discussion attempts a finer-grained characterization of various groups with an interest in a system in order to improve our understanding and conceptualization of various aspects of systems development [Markus & Mao, 2004]. Attention in the academic literature has focused on four main relevant groups in systems development, which are discussed in detail below: developers, users, top management and external agents. The project team forms another, composite, group that is often identified. Finally, consideration is given to the potential influence of interaction between individuals or groups on systems development and its outcomes.

## 4.1 Developers

Software systems development professionals possess a range of characteristics that can influence how they approach and practice development and what contribution they make to a project and its outcomes. These characteristics include: technical skills, capabilities, expertise and experience; interpersonal, communication and social skills; application domain knowledge; commitment, motivation and trustworthiness; and norms, values and beliefs. Significant variation in skills and capabilities of developers can influence development productivity [Fitzgerald, 1998b; Fitzgerald et al., 2002] and hence project outcomes.

Empirical studies suggest that competent staff with adequate technical skills can play an important role in facilitating positive project outcomes [Barry & Lang, 2003; Jiang et al., 1996; Keil et al., 2002; Procaccino et al., 2006; Somers & Nelson, 2001; Standish Group International, 1999]. In particular, developer technical expertise, experience and training are said to have an important influence on project success [Baddoo et al., 2006; Fitzgerald, 1998a; Fitzgerald et al., 2002; Kim & Peterson, 2003; Peterson et al., 2002; Wixom & Watson, 2001]. Conversely, lack of developer expertise and experience is considered to be a project risk and may contribute to poor project outcomes, even project abandonment [Peterson & Kim, 2003; Schmidt et al., 2001].

In a survey of US system project leaders, Aladwani [2002] found that developers' problem-solving competency was perceived to be critical to successful outcomes. This is not surprising as software system projects generally involve identifying and defining problems, generating solutions, reviewing alternatives, and evaluating options. Aladwani also found that clarity of project goals and staff expertise had a positive effect on problem-solving competency, while project team size had a negative effect. According to Fitzgerald et al. [2002], in addition to analytical skills, this process requires creative skills and judgment. An individual's education, training and work background can influence his or her problem-solving approach and ability [Gasson, 1999]. As developers gain more experience, they learn, extending their skill level and building up a repertoire of development strategies [Fitzgerald, 1998b; Fitzgerald et al., 2002]. Aladwani [2002] suggests that since a high proportion of developers' work can involve problems that are very similar in nature, developers with wider experience and knowledge are more likely to have faced similar problems before. In light of rapid advances in technology and changing development practices, Kim & Peterson [2003] suggest that ongoing training may be important to organizations with a continuing commitment to software systems development. Baskerville & Pries-Heje [2004, p. 260] argue that in the short cycle time development characterizing many modern software development projects, "skilled, experienced, and talented developers are needed to anticipate problems and innovate workable shortcuts".

Good interpersonal and communication skills are perceived to be important for interacting with users, and for facilitating dialogue between different groups of users [Baddoo et al., 2006; Fitzgerald et al., 2002; Jiang et al., 1998a; Marion & Marion, 1998; Wixom & Watson, 2001]. Hornik et al. [2003] found low levels of user satisfaction in system development projects where users perceived the developers to have poor communication skills, regardless of their technical expertise.

If development is as much a social and political activity as a technical one, then change management skills may be necessary for developers [Markus & Benjamin, 1996; Symon, 1998]. Howcroft & Wilson [2003] suggest that developers require political skills in order to negotiate the often competing demands or interests of management on the one hand and user groups on the other. Other authors observe that developers may exercise political advocacy and image management skills [Markus & Benjamin, 1996; Symon, 1998]. In the acquisition and use of packaged software, internal professionals are becoming increasingly involved in negotiating contractual and financial issues [Howcroft & Light, 2006].

The outcome of a project can depend on the understanding that development professionals have of the system context or problem domain [Baddoo et al., 2006; Butler, 2003; Fitzgerald, 1998a; Fitzgerald et al., 2002; Sumner, 2000]. This includes their knowledge of organizational operations, sensitivity to organizational norms and politics, understanding of the culture and functioning of user departments, and familiarity with and expertise in the type of application being developed [Jiang & Klein, 2000; Marion & Marion, 1998]. Where developers are outsiders to the organization, they are more likely to have a more limited knowledge of the user and the system's context [Sarkkinen & Karsten, 2005].

Fitzgerald [1998b; Fitzgerald et al., 2002] identified the level of motivation and commitment of developers as an important influence on the outcome of a software systems project. Developers who were motivated and committed to a project were more likely to ensure that the project was successfully completed. According to Oz & Sosik [2000], developer motivation is likely to be influenced by both the organizational context and the composition and culture of the project team. Developers may be motivated by effective leadership, a positive working environment, a sense of being involved, positive feedback, and where they enjoy a reasonable level of autonomy or responsibility. The technical challenge of designing a new solution, the opportunity to work with new technology or standard methods of systems development, working more closely with top management, or being helpful to users may also motivate some developers [Fitzgerald, 1998b; Fitzgerald & Fitzgerald, 1999; Fitzgerald et al., 2002; Jones & Harrison, 1996; Kautz et al., 2004; Linberg, 1999; Nandhakumar & Avison, 1999]. On the other hand, developers may become frustrated or demotivated, not fully committing to the goals of the project, for various reasons, including lack of autonomy, inadequate development resources, technologically unrealistic requirements, conflict between team members or with managers, poor prior development experiences, job insecurity, and use of an inappropriate or constraining standard method or tools [Linberg, 1999; Nandhakumar & Avison, 1999].

Marion & Marion [1998] suggest that, in order to establish a working relationship with others interested in a system, development professionals need to be perceived of as trustworthy and sincere. Based on interviews with experienced Irish developers, Fitzgerald [1998b; Fitzgerald et al., 2002] found that project managers used their knowledge of developers' skills and aptitudes in allocating development work. Within an organization, developers who could be trusted were assigned responsibility for critical tasks or projects. Trust was also an issue where critical roles were assigned to an external consultant or vendor.

Development professionals have norms, values and codes of behavior that influence their conduct, and which reflect their socialization and training, and the professional, social and organizational context in which they work. For example, as a consequence of their professional norms, developers may focus on technical matters at the expense of human or organizational issues [Poulymenakou & Holmes, 1996], which can adversely affect the outcome of a project [Skok & Legge, 2002]. That said, Jiang, Klein, & Balloun [1998b] challenge the assumption that developers can be treated as a homogeneous group with one set of norms and behaviors. Based on a survey of US development professionals (including system analysts, project leaders and system department managers), they found that different developers had different orientations, including technical, socio-political, or user orientations, or various combinations of these. Symon [1998] notes that different developers may contest representations of and rationales for their work.

Developers' values, beliefs and assumptions about the users of a technology, including their roles, abilities and needs, and the context of use, shape a technology such as a software system. Embedded within these beliefs and values is a presumed way of using the technology or system which influences users' interaction with it [Wilson, 2002]. For example, developers may design a system using themselves as typical users, resulting in a system more suitable for expert users [Iivari, 2004a].

According to Wilson [2002], users may become dissatisfied with a system and even resist its use when there is a contradiction between their perceived roles and those roles inscribed for them by the developers of the system. Riley & Smith [1997, p. 309] highlight the problems that may result from developers' "reductive view of a given user situation".

Lyytinen & Hirschheim [1987] conclude that failure by development professionals to appreciate differences in how individuals process information or how they may behave in response to a new system can result in poor fit between the system and its users. They argue that systems development activities tend to assume an idealized, average user, ignoring cognitive, motivational or skill differences within the population of users. The user is often viewed as static, with little or no allowance made for learning or cognitive evolution with the system. This can result in a system in which the user is constrained by technical rigidity of the interface presented to him or her.

### 4.2 Users

A number of authors have highlighted that (perhaps even more so than developers) the users of a system are not a homogeneous group [Butler & Fitzgerald, 1997; Iivari & Igbaria, 1997; Markus & Mao, 2004; Taylor et al., 2002; van Offenbeek & Koopman, 1996] although they may be referred to as such. Rather, 'users' may be made up of groups of individuals from different functional, geographical, vertical and horizontal areas in an organization with potentially different characteristics, interests in a system, and capabilities to influence the course and outcome of a system's development [Asaro, 2000; Markus & Mao, 2004; van Offenbeek & Koopman, 1996].

A review of the extensive literature on users and software systems suggests that users may affect the outcome of systems development in three main ways. Users can have an effect through (1) their expectations of the system being developed; (2) their attitude towards and involvement with the system; and (3) specific characteristics that may affect their ability to utilize the system.

*4.2.1 User expectations*

As a major interested group in any developed system, users' expectations are an important influence shaping a software systems project [Staples et al., 2002]. Lyytinen & Hirschheim [1987] argue that user expectations are value-based beliefs and desires about how the system will serve their interests. While some user expectations are explicitly formulated as system goals and requirements [Lemon et al., 2002], other expectations may remain unarticulated or only vaguely expressed. The latter may be a result of the unclear nature of an expectation, the sheer number and diversity of users, or an inability or lack of opportunity for users to voice their expectations [Lyytinen & Hirschheim, 1987].

Schmidt et al. [2001] note that the growing sophistication of users is leading to higher user expectations of systems.

They also identify the need to manage user expectations so as to avoid the mismatch between user expectations and the system delivered. It is commonly held that this can be achieved by participation of users in systems development, through the creation of more realistic user expectations about the system [Lin & Shao, 2000; Mahmood et al., 2000; McKeen & Guimaraes, 1997; Roberts et al., 2000].

The failure to manage user expectations has been found to be an important risk factor to the successful completion of projects [Keil et al., 1998; Schmidt et al., 2001]. Empirical studies have found that established, managed or realistic user expectations are perceived to be important for system success [Lemon et al., 2002; Mahmood et al., 2000; Somers & Nelson, 2001]. Similarly, unrealistic expectations may inhibit successful systems development [Barry & Lang, 2003].

*4.2.2 User attitude and involvement*

User attitude is usually defined as a psychological state reflecting an evaluative judgment or feeling towards a system [Barki & Hartwick, 1994]. Users' attitudes towards a system affect their intention to use and actual use of the system [Amoako-Gyampah, 1997; Mahmood et al., 2000]. User attitudes may not necessarily result from a reasoned assessment of the functionality of the system. Although nothing can compensate for a lack of needed functionality [Mahmood et al., 2000], often subjective perceptions of the characteristics of a system determine a user's attitude towards it [Amoako-Gyampah, 1997]. Users are likely to have a positive attitude about a system if they perceive that it is useful to them, it is easy to use, or it is in their interests to use it [Mahmood et al., 2000; Wilson & Howcroft, 2002]. Riley & Smith [1997] argue that if users are enthusiastic about a system then other obstacles are less likely to become critical problems.

Negative user attitudes towards a system (which in some cases may lead to resistance to use it) can result from a perceived lack of relevance, consequent changes to the way work is performed, or when users feel threatened by change associated with the new system [Bussen & Myers, 1997]. Empirical studies have shown that the introduction of a system can be problematic in situations where workers have a strong professional culture, identity, autonomy or level of unionization (often evident in the health, education and social care sectors). Problems can arise where the system, or parts of it, are perceived as challenging traditional professional values, roles, status and work conditions, undermining or threatening individual or collective identities, and making work practices more transparent [Doolin, 2004; Marion & Marion, 1998; Myers & Young, 1997; Riley & Smith, 1997; Wilson, 2002; Wilson & Howcroft, 2002].

User involvement (as distinct from user participation, discussed in Section 6.4) is a psychological state that reflects the extent to which a user perceives a system to be both important and personally relevant [Barki & Hartwick, 1994]. Empirical studies have found that user involvement is a significant factor in the successful completion of software systems projects [Hwang & Thorn, 1999]. Conversely, lack of user commitment is considered to be a project risk [Keil et al., 1998; Schmidt et al., 2001] and lack of user support has been found to be negatively related to various measures of project success [Jiang, Chen et al., 2002; Jiang & Klein, 1999, 2000; Jiang, Klein et al., 2000].

Prior work in software systems development has found a relationship between user attitude and user involvement [Hunton & Beeler, 1997]. Changing the attitude of highly involved users tends to require strong persuasive arguments that are factual and logical. In contrast, individuals with low involvement are more likely to change their attitudes because of normative influences such as interpersonal concerns or appeals from others who are important to them. Kirsch & Beath [1996] suggest that the actions of developers (e.g. through education, motivation or negotiation) can increase the extent to which users feel involved in a software systems project.

*4.2.3 Other user characteristics*

User attitude towards and involvement with a system may also be influenced by user characteristics such as personality type, experience with systems and organizational status [Barki & Hartwick, 1994]. Some user characteristics, particularly users' lack of experience with or understanding of systems generally, the specific system or type of application, or the activities the system is intended to support, have been found to have a negative relationship with overall system success and some measures of project effectiveness [Jiang & Klein, 1999, 2000]. Similarly, Mahmood et al. [2000] found that user satisfaction was strongly affected by user background characteristics such as user experience and skills.

**4.3 Top Management**

Top management, either individually or collectively, is often considered to play an important role in systems development, although who exactly comprises this group is generally not defined. In the management literature, top management refers to the group of most senior executives and decision makers with responsibility for the overall strategic direction of the organization [e.g. Wiersema & Bantel, 1992].

The presence or absence of top management support, commitment or understanding continues to be consistently reported in the software systems development literature as important in determining the outcome of a project [Akkermans & van Helden, 2002; Aladwani, 2002; Jiang et al., 1996; Jiang, Klein et al., 2000; Newman & Sabherwal, 1996; Pan et al., 2004; Procaccino et al., 2005; Sharma & Yetton, 2003; Somers & Nelson, 2001]. For example, top management support ranks highly in the Standish Group's CHAOS studies of project success factors, ranking either first or second of ten success factors in 1998 and 2000 [Johnson et al., 2001; Standish Group International, 1999, 2001]. The importance of top management support for software systems success has been observed across a range of national and institutional contexts [Coombs et al., 1999; Kim & Peterson, 2003; Lemon et al., 2002; Peterson et al., 2002]. Several authors suggest that top management support needs to be sustained throughout systems development and implementation if a project is to be successful [Butler & Fitzgerald, 1999b; Keil et al., 1998; Umble et al., 2003].

Similarly, lack of top management support is considered an important project risk factor [Kappelman et al., 2006; Keil et al., 1998; Schmidt et al., 2001; Sumner, 2000] and has been implicated in challenged, abandoned or failed projects [Oz & Sosik, 2000; Martin & Chan, 1996]. In a survey of UK and New Zealand project managers, Yetton et al. [2000] found support for the hypothesis that a project with senior management support was more likely to be completed and not redefined or abandoned.

Top management support may be particularly important in specific systems development contexts, for example, executive information systems (EIS), where the support of an executive sponsor as a potential user is important [Nandhakumar, 1996]; projects that are considered to be strategic or critical to business success [Yetton et al., 2000]; projects that challenge the professional identity or autonomy of users [Riley & Smith, 1997]; projects with high task interdependence [Sharma & Yetton, 2003]; or large systems that have significant impacts throughout the organization, such as customer relationship management (CRM) projects [Kim & Pan, 2006], data warehouse projects [Wixom & Watson, 2001], ERP projects [Mabert et al., 2003; Somers & Nelson, 2001] and manufacturing resource planning projects [Irani et al., 2001].

The importance placed on top management support stems from the various roles that top management is perceived to play in software systems development. For example, top management support is considered important for ensuring the availability of budgetary and human resources required for the project [Aladwani, 2002; Butler & Fitzgerald, 1999b; Kim & Peterson, 2003; Parr & Shanks, 2000; Schmidt et al., 2001; Sharma & Yetton, 2003; Yetton et al., 2000].

Top management is considered to also have an important role in overseeing systems development [Aladwani, 2002; Schmidt et al., 2001; Sharma & Yetton, 2003; Yetton et al., 2000], and ensuring that the project is aligned with and supports organizational strategies and goals [Clegg et al., 1997; Kim & Peterson, 2003]. Failure of top management to monitor progress, support and enforce management and control procedures, or be involved in critical decisions can cause project failure or abandonment [Goldstein, 2005]. According to Standish Group International [2001], top management should be responsible for setting the agenda for the project, and articulating the project's overall goals. It should have an overall understanding of the project and how it benefits the organization. Pan & Flynn [2003] argue that top management has a role to play in managing political conflicts that emerge during systems development and implementation.

Top management support is also considered important in influencing user attitudes, whether actively championing or visibly associating with the project to signal organizational commitment to the project [Parr & Shanks, 2000; Sharma & Yetton, 2003], encouraging user participation in a project [Kim & Pan, 2006; Wilson et al., 1997; Wixom & Watson, 2001], or countering any negative attitudes of users towards the new system or resulting organizational changes [Kim & Peterson, 2003; Riley & Smith, 1997; Yetton et al., 2000]. A significant project may entail the redefinition of roles and responsibilities within an organization. Top management can be influential in creating a positive context for change [Butler & Fitzgerald, 1999b; Lemon et al., 2002; Schmidt et al., 2001; Wixom & Watson, 2001]. According to Sharma & Yetton [2003], top management plays an important role in shaping the organizational context, which can influence how users appropriate a system. They suggest that top management can facilitate successful system implementation by instituting mechanisms or structures that facilitate user learning, instituting performance control systems that recognize and reward use, instituting coordination mechanisms that support the changes associated with a system, and instituting changes to performance goals.

Other influential decision-makers, such as a company's board of directors, may exert a level of influence, particularly in terms of sponsoring a software systems project [Gasson, 1999].

### 4.4 External Agents

According to Sawyer [2001b], the systems development market has changed from the approach of the early 1990s in which organizations largely developed their own systems internally. Increasingly, organizations are sourcing their solutions externally as made-to-order software or ready-to-install software packages. Within this development context, external consultants are playing an increasingly important role, particularly, in bridging the gap between system consumers and software vendors [Howcroft & Light, 2006; Sawyer, 2001b; Skok & Legge, 2002]. External consultants may also be utilized where the organization lacks specific expertise [Butler, 2003], or to 'grow' internal staff expertise [Sumner, 2000]. Although prior studies have found only limited evidence for the importance of the use of external consultants on the outcome of a software systems project [Akkermans & van Helden, 2002; Irani et al., 2001; Schmidt et al., 2001; Somers & Nelson, 2001], with an increasing presence in systems development, their influence on project outcomes can be expected to increase.

Challenges associated with using external consultants or contractors could include the nature of the contract and contractual issues (such as what constitutes an error, enhancement or unforeseen cost) [Goldstein, 2005; Pan et al., 2004]; lack of understanding or misinterpretation of organizational requirements by consultants [Howcroft & Light, 2006; Pan et al., 2004]; lack of control over the actions of external consultants [Schmidt et al., 2001]; poor product quality and poor service [Pan et al., 2004]; communication problems between consultants and users, or no direct communication channels between them [Pan et al., 2004; Skok & Legge, 2002]; high expense [Lemon et al., 2002]; lack of internal systems support once external consultants have departed [Butler, 2003]; and possibly reduced participation of users [Howcroft & Light, 2006; Sawyer, 2001b].

Sarkkinen & Karsten [2005] highlight the difficulties that external developers or consultants can encounter during a project, particularly where the system significantly changes individuals' work practices, task division, and organizational status or authority. As outsiders to the organization, external developers or consultants may be unaware of the consequences associated with the new system or of any political undercurrents. They are more likely to focus on the technical aspects rather than the social aspects of the project. In doing so, they are likely to be perceived by users participating in the project as agents of management, forwarding their interests.

### 4.5 Project Team

All but the smallest of software systems projects are undertaken by a team that includes development personnel, user representatives, managers, and possibly external consultants. The composition of the project team, their collective expertise, their roles and relationships, can influence project outcomes through project team performance. For example, Jiang, Klein et al. [2002] found that strong project team effectiveness improves project outcomes. Similarly, Wang et al. [2005] found that project team cohesiveness was significantly positively related to project performance.

The size and composition of the project team can themselves influence the outcome of a project. For example, large-sized project teams and teams that have not worked together in the past have been suggested as project risk factors [Jiang, Klein et al., 2000]. Aladwani [2002] found that project team size was significantly negatively correlated with project team performance, with larger teams experiencing dissatisfaction among team members and decreased productivity and problem solving. Developers in the team interviewed by Linberg [1999] felt that small-sized teams improved communication, enabled collaboration, and facilitated a sense of synergy. Empirical evidence suggests that a stable, experienced, cohesive project team can lead to good project performance [Yetton et al., 2000]. The most effective development teams are asserted to be those with a balance of diverse personality types and mutual openness to ideas [Bradley & Hebert, 1997; Linberg, 1999].

Project team skills have also been found to have a major influence on project outcomes. According to Aladwani [2002], a project team with a variety of experience and skills is likely to perform better than one with lesser available skills. It has been suggested that for effective project team functioning, the collective expertise of the project team should enable them to accomplish the range of allocated tasks, to work with undefined elements, uncertain objectives and issues emerging during the project, to work cooperatively as a team and with top management, and to understand organizational operations and the human implications of the system [Jiang & Klein, 2000; Jiang, Klein et al., 2000; Kim & Peterson, 2003; Wixom & Watson, 2001].

In a survey of data warehousing managers and users, Wixom & Watson [2001] found that a project team with strong technical and interpersonal skills was able to perform tasks well and interact with users, leading to project implementation success. A skilled and competent project team was considered to be more able to identify the complex project requirements. Wixom & Watson [2001] concluded that in projects that involve specialized technology it is important that the development team understand how to use the technology and how it relates to the existing technical infrastructure. Jiang, Klein et al. [2000] suggest that where teams lack sufficient expertise with the application or technology being developed, they may become reliant on the few team members who do, leading to inefficient use of team resources. In three case studies of CRM system implementations, Kim & Pan [2006] found that the balance between high levels of business skills and technical expertise within the project team in a successful implementation was missing in the unsuccessful cases.

The use of support technologies and tools can supplement the capabilities and productivity of the team [Aladwani, 2000, 2002]. However, in a study of software project teams performing requirements analysis, Guinan et al. [1998] found that group processes and team performance were positively influenced more by project team skill, the project manager's involvement in the day-to-day workings of the team, and similar levels of experience within the team, than by the use of systems development methods and tools. Similarly, Sawyer & Guinan [1998] found that the use of automated development tools had no explanatory effect on variances in either software product quality or project team performance.

The roles and responsibilities of the various team members need to be well-defined and clearly communicated to team members. Improper definition of roles and responsibilities has been perceived as a risk to successful completion by both project managers and system users [Keil et al., 2002; Schmidt et al., 2001]. Empirical studies have found that lack of clarity of role definition is significantly negatively related to system success [Jiang & Klein, 1999, 2000]. When roles and responsibilities are poorly defined or communicated, requirements may be overlooked, items or features may be left out or not completed, or there may be significant task overlap [Keil et al., 2002].

### 4.6 Social Interaction

The development of a software system can be perceived (though not exclusively) as a social process involving interaction between participants in various social roles [Kirsch & Beath, 1996]. Throughout systems development, individuals from the groups described above may interact in various ways, including negotiation, decision-making, communication, conflict or political manoeuvring. This interaction will be shaped by similarities and differences in the various groups' values and beliefs, professional or social norms, expectations and perceived interests.

Individuals who are perceived by other participants in a project to be experts in some area (for example, with knowledge of systems development practice or of the application domain) can shape the meaning of systems development (and its activities) for others [Gasson, 1999, 2006; Symon, 1998]. Gasson [1999] found that individual experts managed meanings to the extent that they defined what were appropriate forms of the systems development process, its products, work roles and activities. She suggests that such influence may diminish as other areas of knowledge become more important in a project.

The nature and quality of interactions between participants, particularly users and developers, can influence project outcomes [Procaccino et al., 2006; Robey & Newman, 1996; Wang et al., 2006]. Changes in the relative influence of groups, and critical encounters between them, can affect the course of a project [Heiskanen et al., 2000; Robey & Newman, 1996]. It has been suggested that the key to establishing a working relationship between project participants is the creation of mutual respect and trust – a responsibility that often falls to development professionals or to the project manager [Marion & Marion, 1998]. A shared organizational culture can also be a basis for positive interaction [Butler & Fitzgerald, 2001; Poulymenakou & Holmes, 1996; Symon, 1998].

There are often multiple direct and indirect channels for interaction among and between system participants. In bespoke developed projects, these may include facilitated workshops, intermediaries, a customer support line, prototyping, interviewing, testing, a survey, email or a bulletin board, and observation of work tasks. User participation may be a means for developing a social relationship between users and developers [Kirsch & Beath, 1996]. According to Fitzgerald et al. [2002], during systems development users and developers learn from each other in a mutual, interactive way. The use of standard methods of systems development can influence interactions between participants in a software systems project by structuring roles, responsibilities and occasions for interaction [Robey et al., 2001].

Asaro [2000] suggests that in situations where the emerging system artifact becomes part of the system development, it mediates user-developer interaction. Developers cannot interpret requirements in isolation of users' reactions to the developing system, and users can less easily resist a system that has been built and revised in response to their concerns. Both groups also become aware of the practical and material limitations of the technology itself. For example, Hardgrave et al. [1999] suggest that prototyping facilitates increased and more responsive interaction and communication between users and developers. Butler & Fitzgerald [1997; 1999a; 1999b; 2001] describe how in certain projects the use of prototyping or computer-aided software engineering (CASE) tools improved user-developer communication, and increased the level of user participation and involvement in the projects by providing a common language that bridged the traditional gap between technically-oriented developers and business-oriented users.

Effective interaction between participants in a software system project can facilitate the alignment of goals and expectations, achieve mutual understanding, and encourage effective communication. However, it can also lead to more contradictory outcomes when differences between participants emerge, or when misunderstandings or breakdowns in communication occur.

### 4.6.1 Goals and expectations

The recognition that there are typically multiple interested participants in a project, each with different interests, values, beliefs, norms, practices and behaviors, rewards, goals or expectations has led some authors to argue that successful software systems development relies on alignment or congruence between such attributes [Jiang, Chen et al., 2002; Jiang, Sobol et al., 2000; Keil et al., 2002; Marion & Marion, 1998; Pan, 2005]. For example, Jiang, Chen et al. [2002] suggest that different groups of participants will have different expectations and will judge the system being developed on different criteria.

Substantial differences in goals and expectations can occur between groups of development professionals, between system developers and the users of a system, or between different groups of users. For example, Mahaney & Lederer [2003] found a perceived goal conflict between system developers and project managers, with respect to solution quality and delivery, respectively. In a study of US systems professionals and users reported by Jiang, Sobol et al. [2000], development personnel often believed that they had reached agreement with users over project objectives, whereas the users did not believe such an agreement had been reached. Consequently, the users, who had different expectations of the system, were dissatisfied when it failed to meet their expectations.

Different groups of users may potentially have conflicting organizational interests or professional interests [Doolin, 2004; Marion & Marion, 1998; Myers & Young, 1997; Riley & Smith, 1997; Wilson & Howcroft, 2002]. Further, the interests or expectations of participants are not necessarily static and may change over the course of systems development [Pan, 2005]. This may occur, for example, through the development of coalitions of individuals, or as members of the project team develop loyalty to each other and/or the project [Myers & Young, 1997].

Jiang, Chen et al.'s [2002] solution to goal or expectation incongruence is a compromise between the various groups in order to reconcile their differences. They view project management as the exercise of this compromise, in the face of resource constraints and the realization that no one set of needs will be completely satisfied. As part of this stance, Jiang, Klein et al. [2000] argue that the common interests of various stakeholders should be emphasized. Jiang, Chen et al. [2002] suggest that pre-project partnering, in which various stakeholders work together before a project begins, is a useful approach for fostering collaboration and reducing the potential for conflict. Surveying US development professionals, Jiang and co-authors found that pre-project partnering was significantly positively associated with project performance. They also noted that pre-project partnering reduced the risk of poor user support for the project, and led to effective project team characteristics and improved project manager performance [Jiang, Chen et al., 2002; Jiang et al., 2006; Jiang, Klein et al., 2002].

### 4.6.2 Understanding

Historically, a lack of understanding between participants in a software system project has been associated with system failure [Sauer, 1999]. Based on a survey of systems development participants, Enquist & Makrygiannis [1998] found that misunderstandings occur frequently between them throughout the development process. Such misunderstanding often produces minor negative consequences (such as minor process delays, product errors, and/or problems in relations with other participants), but occasionally their consequences may be more extensive. The most common causes of misunderstandings were (in order) unclear or incompletely expressed information; differences in concepts and frames of reference; and uncertainty about tasks, responsibility, authority or intentions of other participants.

A gap in understanding (in particular, between users and developers) has been attributed to differences in organizational cultures or sub-cultures [Al-Karaghouli et al., 2005; Coughlan et al., 2003; Enquist & Makrygiannis, 1998; Flynn & Jazi, 1998; Jiang, Sobol et al., 2000; Poulymenakou & Holmes, 1996; Taylor-Cummings, 1998]. Such cultural divergence can arise from differences in organizational roles and loyalties, professional backgrounds, world views, interests, expectations, skills bases, experience, ambitions, education, training, cognitive styles, problem-solving approaches and vocabularies [Butler & Fitzgerald, 1997; Flynn & Jazi, 1998; Gasson, 1999; Jiang, Sobol et al., 2000; Symon, 1998; Urquhart, 2001]. A common perception is that developers are focused on technical issues, while users are concerned more with facilitating work or business tasks. For example, based on a survey of developers and users in the UK, Al-Karaghouli et al. [2005] attribute the gap in understanding (rather unsurprisingly) to lack of business knowledge by developers and lack of technical understanding by users.

Differences in understanding can also be viewed as a result of the diverse interpretive schemes or frames used by various participants to construct meaning in relation to the project [Galliers & Swan, 2000]. For example, individuals (with different education, training, work background and prior experiences with systems development) may have different perceptions (and preconceptions) of the purpose, meaning and use of a system, which may influence their ability to achieve a shared understanding of the new system [Gasson, 1999]. In a case study of software system design, Gasson [1999] observed that individual project team members influenced each others' perspectives on the project, and that these perspectives converged with time as the team developed a shared understanding of the project.

### 4.6.3 Communication

Communication is often perceived to be an important dimension of the interaction between users and development staff, essential for effective functioning of the project team, and a key factor in system success [Akkermans & van Helden, 2002; Butler, 2003; Butler & Fitzgerald, 2001; Hartwick & Barki, 2001; Sawyer & Guinan, 1998; Somers & Nelson, 2001]. Conversely, poor communication can lead to misunderstanding and conflict between participants, which may even be carried over into subsequent projects within the organization [Amoako-Gyampah & White, 1997; McKeen & Guimaraes, 1997; Skok & Legge, 2002].

Communication between participants in a project can be informal or formal, direct or indirect, one-way or two-way [Amoako-Gyampah & White, 1997; Butler, 2003; Butler & Fitzgerald, 2001; Gallivan & Keil, 2003]. Communication is influential through the role it plays in facilitating information exchange, mutual understanding and collaboration, and in identifying and resolving conflicts [Amoako-Gyampah & White, 1997; Keil et al., 2002; Oz & Sosik, 2000]. It has been suggested that establishing a shared language or vocabulary between participants is important for achieving effective dialogue [Marion & Marion, 1998].

Effective communication is frequently perceived as important for meaningful user participation in software systems projects [Amoako-Gyampah & White, 1997; Hartwick & Barki, 2001]. It is considered necessary for users to convey their understandings of the organizational context and their requirements to developers, and for developers to explain technical issues to users and listen to user-related problems [Al-Karaghouli et al., 2005; Butler & Fitzgerald, 2001]. However, Gallivan & Keil [2003] suggest that 'communication lapses' may occur that negate or reduce the effectiveness of user participation. Such communication lapses can occur where development is framed in a way that excludes consideration of particular issues; where users are unaware of an issue being a problem, see no need to communicate an obvious problem, or are unable to articulate an issue as a problem; where user representatives may not perceive an issue as problematic even though other users might; where communication channels are not available or where users are unaware of those communication channels available; where users actively decide not to communicate through a channel because certain messages are perceived as politically or socially unacceptable; where interpretive schemes, mental models, differences in language use, or intermediaries distort or filter out specific messages; or where developers fail to act on a message, act on the wrong messages or consider certain actions unacceptable [Gallivan & Keil, 2003].

Communication may also be used by the project team as an important component in maintaining relationships with, and the support of, other organizational groups [Jiang, Klein et al., 2000]. Amoako-Gyampah & White [1997] note the need for ongoing two-way communication so that users and managers feel that their input is valued (and will be sought), are given feedback on their input or concerns, and are informed about project changes. In a post-hoc longitudinal case study, Butler & Fitzgerald [1999b] found that the project manager, developers and users had employed various strategies

(such as a high degree of formal and informal communication between groups) to avoid 'us vs. them' scenarios developing.

### 4.6.4 Conflict and politics

Differences in values, perceptions, interests, goals or expectations, a lack of mutual understanding, and ineffective communication, have all been attributed to causing disagreement or conflict between participants in a project. Conflict may occur between groups associated with systems development, such as users or developers, and within such groups [Symon, 1998], including the project team. Coakes & Coakes [2000] suggest that conflict can arise between different groups or individuals with apparently similar interests because of different interpretations of a problem. Conflict may also be of an interpersonal nature. For example, conflicting personalities and attitudes may lead to poor project team relationships [Keil et al., 2002; Schmidt et al., 2001].

Unsurprisingly, the presence and intensity of conflict and disagreement between participants can adversely impact the systems development process and project outcomes [Jiang & Klein, 2000; Keil et al., 2002; Pan, 2005; Robey & Newman, 1996; Schmidt et al., 2001]. Poor relationships between participants may continue until they are disrupted by conditions that challenge existing behavior [Robey & Newman, 1996].

The literature suggests that the potential for conflict increases as the number and diversity of participants involved increases, when the scope of the project is large, when the project is highly complex, when high levels of integration among the participants is necessary, and when external factors such as third parties or other projects are involved [Linberg, 1999; van Offenbeek & Koopman, 1996; Yetton et al., 2000]. Robey et al. [2001] argue that conflict increases as role interdependence between participants increases, especially under time or resource constraints and when responsibilities or approaches to the work differ. Developers interviewed by Linberg [1999] indicated that conflicts often occurred both within the project team and with external managers, sometimes as a result of the pressure under which developers were working. Similarly, Sawyer [2001a, p. 174] sees conflict as inevitable when people interact in activities such as systems development, which are "characterized by ambiguity, contradictory information and time pressures".

Robey et al. [2001] suggest that conflict can sometimes have a positive effect if it encourages meaningful and constructive debate among participants. Acknowledgment of disagreement and conflict can ensure that important project issues are addressed and new or creative solutions are considered [Wilson et al., 1997], arguably leading to better decision making [Sawyer, 2001a]. The participation of various groups in systems development has been suggested as a way of reducing potential conflict in projects. The rationale for this position is the increased level of mutual understanding between different groups through their working together [Jiang, Chen et al., 2002] or the increased sense of ownership and control engendered through their involvement [Butler & Fitzgerald, 1999b]. However, conflict resolution may not always be achieved through the articulation of differences and the negotiation of a shared understanding or compromise [Jiang, Chen et al., 2002; Sawyer, 2001a].

Given the long term consequences of what is at stake, it is not surprising that the level of political activity in systems development can be high [Butler, 2003; Clegg et al., 1997; Foster & Franz, 1999; Howcroft & Wilson, 2003; Myers & Young, 1997]. A number of studies have found that in certain cases organizational politics can adversely affect the outcome of a project [Robey & Newman, 1996; Warne & Hart, 1996; Yetton et al., 2000]. Pan & Flynn [2003] identified a number of political issues that influenced decision making or produced conflict in an electronic commerce system project, leading to its abandonment. These were political mistrust among project stakeholders (including those external to the organization), formation of an opposing coalition, threats of retaliation, political insensitivity, lack of political promotion of the system project, and failure to obtain continued political support from top management. Politics may become a problem with organization-wide systems (or even industry-wide systems) that span multiple groups who feel their interests (e.g. their ownership and control of business processes and data) are being threatened enough that they decide to take action [Drummond, 1996; Gasson, 2006; Warne & Hart, 1996]. Akkermans & van Helden [2002] found that open communication and a lack of political behavior among different organizational groups were important in turning around a failing ERP project.

Participants may draw on prevailing norms, values and resources to legitimize their actions (e.g. to justify using a particular development approach or method or to include or exclude various groups or individuals from participating) or to mask their political motives [Fitzgerald, 1998b; Fitzgerald et al., 2002; Howcroft & Wilson, 2003]. Butler & Fitzgerald [2001] describe how in a case of shared project ownership, different user groups resorted to political infighting in order to influence the development team in their favor. Butler [2003] also describes how friction developed between two user business units in a corporate software systems project. Doolin [1999] describes a political struggle for control of a systems project, but between the development department and a competing source of authority within the organization, the Finance department, which contested the perceived validity of the system solution.

Myers & Young [1997] describe how, in a software systems project in the health sector, user participation was used to legitimize the project amongst the wider user community. Further, senior management had a hidden agenda. Features that clinical users perceived as challenging their professional status were omitted from the initial user requirements, and were not discussed until the project was well underway and the project team and user representatives had built up allegiance to the project.

Developers themselves, often lacking formal organizational authority, may also use political tactics to secure access to necessary resources, to work around management-imposed constraints, or to secure the support and cooperation of other organizational groups [Linberg, 1999; Nandhakumar, 1996]. Alternatively, systems professionals may fail to support an software systems project that is not under their control [Olesen & Myers, 1999].

Key properties or attributes of the influences that comprise the people and action dimension are summarized in Figure 3.

**Figure 3:** Properties/attributes of influences related to people and action

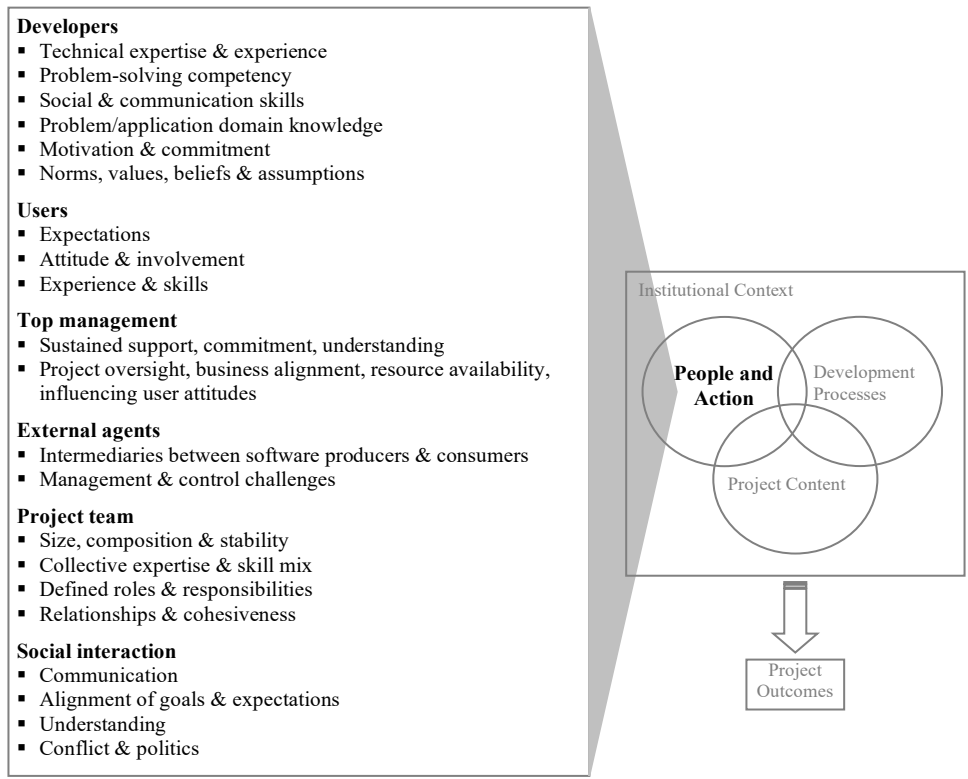

## 5. INFLUENTIAL FACTORS – PROJECT CONTENT

This section discusses influences related specifically to software systems development and deployment projects, including the characteristics of the project, its goals and objectives, the resources made available to it, and aspects of technology that support the system and its development.

### 5.1 Project Characteristics

There is some evidence that the outcome of a software systems project is related to various characteristics of the project itself, such as its size, technical complexity, and newness to the organization [Jiang & Klein, 1999; Johnson et al., 2001; Martin & Chan, 1996; Standish Group International, 1999; Yetton et al., 2000].

Project size may be characterized (if not measured) in a number of ways, including the duration of the project, its cost, the size of the project team, the number of different stakeholders on the team, the number of users, the number of organizational units involved, and the number of hierarchical levels occupied by system users [Jiang & Klein, 2000; Standish Group International, 1999]. Each indicator provides some (limited) sense of the scale of the project, the implication being that higher values generally indicate larger and therefore more difficult projects to be undertaken. Johnson et al. [2001] attribute part of the increase in project success rates observed in the USA between 1994 and 2000 by the Standish Group to smaller project size. They suggest that the emergence of the World Wide Web and the use of standard software infrastructures have facilitated the development of smaller-sized projects, which has in turn led to proportionally more successes. Some organizations have dealt with large projects by breaking them down into smaller sub-projects or by using incremental development [Kautz et al., 2004; Pan et al., 2004] in an effort to increase the likelihood of a successful outcome.

Chatzoglou [1997] suggests that smaller projects tend to involve well-defined project domains, facilitating more effective determination and implementation of system requirements. Large projects are more likely to have high complexity and high task interdependence, need to be redefined, take longer to complete, require more resources, and involve increased lines of communication and potential conflict [Schmidt et al., 2001; Yetton et al., 2000]. In their survey of UK and New Zealand project managers, Yetton et al. [2000] found that project size was negatively related to project completion.

As software systems have become pervasive in organizations, increasing functionality, scope and speed of technical change, and the need for systems integration, have contributed to an increase in complexity [Clegg et al., 1997]. A number of authors have suggested that technical complexity adversely affects project outcomes, including aspects such as project completion and delivery of expected benefits [Barry & Lang, 2003; Jiang & Klein, 2000; Parr & Shanks, 2000]. Jiang & Klein [1999] found a significant negative relationship between application complexity and overall system success. High project complexity can pose problems for various people associated with a software system. Lyytinen & Hirschheim [1987] suggest that, in such cases, individuals often find it difficult to understand the system, and to articulate and act on their concerns.

A project that is new to an organization (in terms of application domain, required functionality) can pose problems because the organization may lack the relevant knowledge, skills or competencies to successfully complete it. To access these skills or competencies, the organization may outsource (part of) the project [Yetton et al., 2000], introducing further risk. Yetton et al.'s survey of project managers [2000] found that newness reduced the chance of project completion, with newer projects being more problematic and more likely to be redefined.

### 5.2 Project Scope, Goals and Objectives

A number of studies have highlighted the importance to project success of an appropriate and achievable project scope, and well-defined and clear project goals or objectives [Aladwani, 2002; Jiang et al., 1996; Kim & Peterson, 2003; Martin & Chan, 1996; Peterson et al., 2002; Somers & Nelson, 2001]. Empirical findings suggest that less than successful project outcomes can arise from excessively large project scope, underestimating the scope of a project, changing scope or objectives, unclear goals or objectives, lack of agreement on goals or objectives among interested parties (e.g. management, information systems staff, users), or elusive goals that emerge and change as the project proceeds [Barry & Lang, 2003; Keil et al., 1998; Keil et al., 2002; Oz & Sosik, 2000; Pan et al., 2004; Parr & Shanks, 2000; Schmidt et al., 2001].

It has been argued that clear project goals can help software system projects address the needs and expectations of both users and the organization. In this sense, the project goals actively guide the determination of information requirements [Kim & Peterson, 2003]. According to Aladwani [2002], clear, well-defined project goals enable the project team to develop a common understanding of the problem and so develop a unified (and therefore less risky) approach to solving it.

Yetton et al. [2000] point out that success is more likely when project goals are well communicated to all concerned with the project. They view clarifying and communicating project goals or objectives to be the role of senior management. Aligning project goals with the goals of the organization is also perceived to be important in ensuring that the system as delivered supports organizational strategies [Aladwani, 2002; Clegg et al., 1997; Kim & Peterson, 2003; Martin & Chan, 1996; Peterson et al., 2002; Poulymenakou & Holmes, 1996]. Clegg et al. [1997] found that the integration of technology and business goals was regarded as the responsibility of senior management.

These challenges are enduring. According to Lyytinen & Hirschheim [1987], part of the cause of systems development failure lies in the fact that goals are often ambiguous, particularly with respect to technical, data, user or organizational requirements. Project goals tend to focus on quantitative aspects, such as the technical aspects of systems development and the economic aspects of organizational performance. Furthermore, goals reflect values – often those of management or information systems professionals – that may later be incorporated into the system being developed. The uncritical adoption of tangential goals and narrow perspectives can lead to 'expectation failure', particularly on the part of users.

### 5.3 Resources

Like any project within an organization, the level of resources made available to a software systems project (including money, people and time for development and implementation) can be central to its outcome (Table 3). Not only is the provision of adequate resources perceived to be important for ensuring successful systems development, but the allocation of inadequate resources is often perceived as contributing to the problems encountered in challenged or failed projects. Even where adequate resources are made available, problems can arise where the project exceeds its allocated costs or project schedule or where the project schedule is altered in some way [Linberg, 1999]. Provision of adequate resources can be particularly critical to organization-wide systems, which can be expensive, time-consuming and resource-intensive [Wixom & Watson, 2001].

**Table 3:** Contribution of resources to system project outcomes

| |
|---|
| **Financial resources** |
| • Adequate financial resources perceived to be important to successful systems development [Fitzgerald, 1998a; Jiang et al., 1996; Martin & Chan, 1996; Nandhakumar, 1996; Wixom & Watson, 2001] |
| • Inadequate financial resources perceived as contributing to problems encountered in software systems projects [Jiang et al., 1998a] |
| **Development time** |
| • Adequate development time perceived to be important to successful development [Fitzgerald, 1998a; Martin & Chan, 1996; Wixom & Watson, 2001] |
| • Inadequate development time or unrealistic deadlines perceived as contributing to the problems encountered in projects [Jiang et al., 1998a; Linberg, 1999; Oz & Sosik, 2000; Schmidt et al., 2001] |
| **Human resources** |
| • Adequate or appropriate project staff perceived to be important to successful development [Jiang et al., 1996; Martin & Chan, 1996; Wixom & Watson, 2001] |
| • Insufficient or inappropriate project staff perceived as contributing to problems encountered in projects [Barry & Lang, 2003; Jiang et al., 1998a; Keil et al., 2002; Linberg, 1999; Nandhakumar, 1996; Schmidt et al., 2001] |
| • Project staff turnover perceived as contributing to problems encountered in projects [Bussen & Myers, 1997; Schmidt et al., 2001; Sumner, 2000; Yetton et al., 2000] |

The allocation of adequate resources can indicate senior management support and commitment to a project, can help to overcome organizational obstacles, and can enable the project team to meet project milestones [Wixom & Watson, 2001]. On the other hand, perceived unwillingness of the organization to provide adequate resources can demotivate members of the project team, causing them to question the project's importance and to not fully commit to the project. Furthermore, unrealistic project schedules can result in extreme workload pressures that undermine developer creativity and compromise project quality [Linberg, 1999].

The effect of human resources is not confined solely to insufficient staff numbers for development. People with appropriate technical infrastructure skills are also needed [Schmidt et al., 2001]. Limited access to technical expertise in certain areas or competition between projects for common human resources can adversely affect or delay a project [Linberg, 1999; Nandhakumar, 1996]. Project staff turnover, especially the loss of key project personnel, can remove critical knowledge about the new system, causing time delays and a loss in user confidence that the system will meet specifications [Schmidt et al., 2001].

### 5.4 Technology

There are clearly many considerations relating to either hardware or software that can potentially influence the outcome of a software systems project. Inappropriate technology selection or use, rapidly changing or new technology, inadequate or inappropriate technical resources available to design and build a system, and difficulties with data, can all result in a challenged technological solution [Kim & Peterson, 2003]. The level of software modification undertaken can negatively impact on project success in packaged software projects such as ERP implementations [Mabert et al., 2003; Sumner, 2000].

The use of appropriate technology is perceived to be important for system success in some cases [Kim & Peterson, 2003; Nandhakumar, 1996; Peterson et al., 2002; Somers & Nelson, 2001; Wixom & Watson, 2001], but not necessarily in others [Jiang, Klein et al., 2000; Oz & Sosik, 2000; Yetton et al., 2000]. It may be that there is an indirect effect – for example, the increasing software and hardware options available means that the technology infrastructure of an organization and the technical expertise available are important considerations in whether or not particular technologies are appropriate [Kim & Peterson, 2003]. Because a high proportion of application code is infrastructure (70% on average), it has been suggested that purchasing standard software infrastructure rather than building it can positively influence project outcomes [Johnson et al., 2001]. The use of an appropriate technical architecture can be helpful for managing project complexity [Vidgen et al., 2004].

Wixom & Watson [2001] found that the use of appropriate 'development technology' (comprising the hardware, software, methods and tools required to complete a project) was significantly associated with successful technical implementation. They suggest that development technology influences the efficiency and effectiveness of the project team. Aladwani [2000; 2002] found that adequacy of development tools was significantly positively associated with project performance. Not only is use of appropriate hardware and software technologies important for delivering an adequate technological solution, it can be important for ensuring user acceptance. For example, in an in-depth study of an EIS development, the use of impressive interfaces was perceived to be important in ensuring executive acceptance [Nandhakumar, 1996].

The introduction of unproven or new technology is also perceived by some to be an important risk factor in various aspects of successful completion of a software project [Jiang & Klein, 1999; Keil et al., 1998; Schmidt et al., 2001; Wastell & Newman, 1996], although Jiang & Klein [2000] reported no significant relationship between technological newness and project effectiveness. However, Wastell & Newman [1996] identified the use of proven software as a critical factor in a case study of successful systems development.

The impact of various technical problems that may arise during the course of a project on the outcome of that project can be mediated by the technical expertise available. For example, in a case study of four projects reported by Butler & Fitzgerald [1999b], various technical problems were encountered with introducing client-server architectures, developing a corporate data warehouse, evaluating hardware platforms, and integrating and interfacing new and existing systems. Overcoming project technical obstacles was perceived to be critical to the success of the development process, and required significant developer or vendor technical skills and expertise.

Data can also present problems to a software systems project. In designing and developing a new solution, the data may be incorrect or in an inappropriate form [Bussen & Myers, 1997; Nandhakumar, 1996]. Inadequate management of data issues, such as availability, ownership and security, can also lead to problems for the project team, such as lack of cooperation from groups outside the scope of the project [Nandhakumar, 1996]. Data quality is particularly critical in the development and implementation of enterprise-wide systems, given the need for data integration across the organization [Somers & Nelson, 2001; Sumner, 2000; Umble et al., 2003; Wixom & Watson, 2001].

Key properties or attributes of the influences that comprise the project content dimension are summarized in Figure 4.

## 6. INFLUENTIAL FACTORS – SYSTEMS DEVELOPMENT PROCESSES

This section discusses influences related to aspects of the systems development process. In particular, it deals with processes of requirements determination, project management, use of a standard method, user participation in the systems development process, user training, and the management of change arising from systems development and implementation.

## 6.1 Requirements Determination

Requirements determination is widely regarded as a critical step in software systems development [Alvarez, 2002; Coughlan et al., 2003; Flynn & Jazi, 1998; Urquhart, 1999, 2001]. Essentially, requirements determination involves achieving a shared understanding of the information, processes and functions that need to be incorporated into the new system [Al-Karaghouli et al., 2005; Coughlan et al., 2003; Urquhart, 1999, 2001]. Although there are often many individuals or groups with an interest in a system, expectations and functional needs are often elicited from the intended users of the system [Lemon et al., 2002]. In addition to user requirements, there may be business requirements that the system will need to satisfy, or technical requirements related to the existing IT infrastructure, the need for integration with other systems, regulatory requirements, or the system itself in the case of packaged software acquisition. A (formal) requirements specification document is usually produced that specifies what the system should do, and often functions as a contract between the project team and the sponsors of the system. It can also serve to guide subsequent design activities. The realization of user requirements – delivering a system that matches the users' needs – is perceived as important by various parties with an interest in a system, including development managers and staff, and users and their managers [Li, 1997].

**Figure 4:** Properties/attributes of influences related to project content

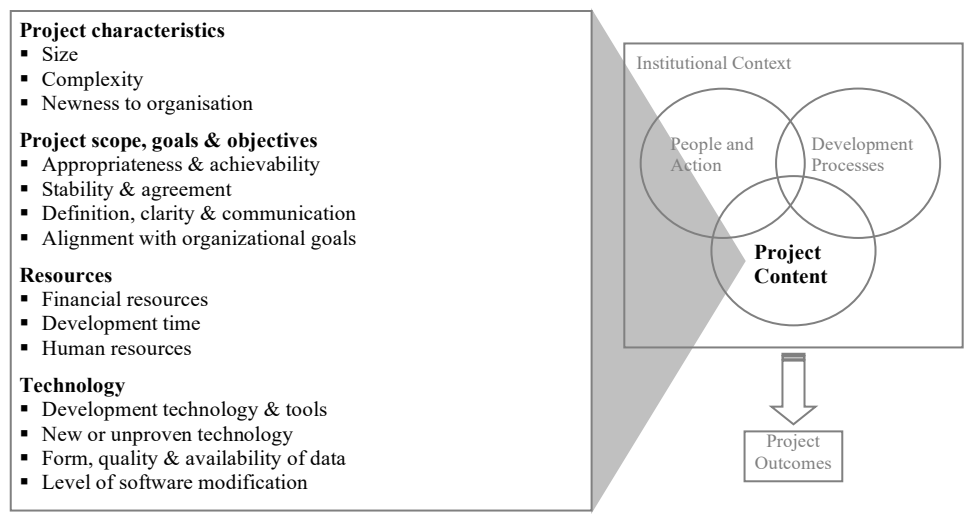

A number of empirical studies have highlighted the importance of well-defined and clearly stated requirements to project success [Lemon et al., 2002; Procaccino et al., 2005; Procaccino et al., 2006; Verner & Evanco, 2005]. Similarly, a lack of or misunderstood requirements is considered to be a project risk factor [Kappelman et al., 2006; Keil et al., 1998; Keil et al., 2002; Schmidt et al., 2001]. Other authors observe that poorly defined or unclear requirements are often an important factor in challenged or abandoned projects [Barry & Lang, 2003; Bussen & Myers, 1997]. Unstable or changing requirements are also perceived to render software projects problematic [Barry & Lang, 2003; Keil et al., 1998; Oz & Sosik, 2000; Schmidt et al., 2001]. Strategies that have been suggested to counter the risk of changing requirements include using iterative design, in which different parts of the system's functionality are delivered in different phases [Johnson et al., 2001; Larman & Basili, 2003; Schmidt et al., 2001], prototyping [Beynon-Davies et al., 1999; Hardgrave et al., 1999] or agile methods [Vinekar et al., 2006; Williams & Cockburn, 2003].

Software projects in which the system requirements are poorly defined can experience difficulties because the resources required in order to complete the project are not fully understood or made available [Butler & Fitzgerald, 1999b; Schmidt et al., 2001]. Poor requirements determination can also result in unclear objectives for the project team, or a system that does not meet the needs and expectations of one of the groups with an interest in it. The latter can result from a failure to identify and include in the determination of requirements all parties with an interest in a system [Pan et al., 2004; Schmidt et al., 2001]. This is particularly relevant where the system spans multiple, diverse groups whose needs must be understood and communicated to the project team [Wixom & Watson, 2001]. The beliefs, ideas and assumptions held by those developing the system can influence the conduct of requirements determination, particularly in terms of who is (and is not) included.

Drawing on a post-hoc longitudinal case study of four systems development projects, Butler & Fitzgerald [1999b] argue that, for system project success, adequate time needs to be spent with relevant users to elicit user requirements. They observed that the outcome of requirements determination depends both on the time that is allocated to it and, more importantly, on the participation of suitable users. Butler & Fitzgerald [1999b] note that within their case study organization, decisions concerning the time allocated for requirements determination were not usually made by the development team, and were often made in response to external conditions, without due regard to what requirements determination actually entailed.

Requirements determination is a complex social process. Various authors have highlighted the importance of communication and mutual understanding between participants with particular reference to the construction of system requirements [Al-Karaghouli et al., 2005; Coughlan et al., 2003; Flynn & Jazi, 1998; Guinan et al., 1998; Urquhart, 1999, 2001]. Problems in requirements determination can arise because users may be unable or unwilling to articulate their requirements, or they may not even know them. Different user constituencies may have different requirements or differing viewpoints on requirements. Users and developers often speak different 'languages' and have different frames of reference. Even if users are willing or able to share their requirements, these are typically translated for design and implementation by developers in most systems development approaches. Users may utilize different mental models or ontological views of organizations and systems. They may not understand (or support) requirements models used by developers or technically-oriented modeling languages. Developers may not sufficiently understand users' work or needs, or may be unable to elicit user requirements, or may think they know already what is required. Further, they may have interests or objectives that take precedence over meeting user requirements (e.g. maintaining technical credibility or technical design integrity) [Al-Karaghouli et al., 2005; Alvarez, 2002; Flynn & Jazi, 1998; Guinan et al., 1998; Urquhart, 2001].

These problems may be compounded by many approaches to (and tools used in) requirements determination, which tend to assume that requirements are objective artifacts that can be codified, specified at the outset, and remain unchanged during development. Such approaches may not adequately recognize the emergent and socially-constructed nature of requirements, nor the political aspects of requirements determination, in which system stakeholders may have different goals, objectives and interests. Often insufficient attention is paid to the social and political context in which the system will be situated [Flynn & Jazi, 1998; Galliers & Swan, 2000].

## 6.2 Project Management

In general terms, software systems development project management involves planning, organizing and managing organizational resources, both financial and human, for the duration of a project. Given the complex nature of such projects, the intricacy of the social interactions that can occur in and around systems development, and the dynamic nature of the development context, it is hardly surprising that empirical studies have emphasized the perceived value placed on project management by the various parties involved in a systems project [Butler & Fitzgerald, 1999b; Jiang et al., 1996; Lemon et al., 2002; Linberg, 1999]. According to Johnson et al. [2001], the increased project success rate observed between 1994 and 2000 in the Standish Group's CHAOS studies was due in part to improved project management processes, better management tools, and more highly skilled project managers. An international survey in 2005 found that the organizational profile of project management continues to increase, with more organizations using project management processes, having project management offices (PMOs), using business cases to justify investment in systems projects, and undertaking project governance for selecting and approving projects (but less so for monitoring projects and measuring benefits) [KPMG, 2005].

Project planning activities include defining the project; estimating its size, cost, and scheduling; assessing potential risks; and developing a project plan. Such activities are usually undertaken by the project manager or leader, a steering committee or an *ad hoc* planning group. Empirical studies have highlighted the perceived importance of planning activities to successful project outcomes [Aladwani, 2000; Barki et al., 2001; Butler & Fitzgerald, 1999b; Kim & Peterson, 2003; Lemon et al., 2002; Mabert et al., 2003; Peterson et al., 2002]. In their survey of project managers, Yetton et al. [2000] found that project planning reduced budget variances, but had no effect on project completion rates. Planning was also found to reduce project team instability. Consistent with these findings, inadequate or insufficient planning, poor estimates, and poor risk management have been held responsible for detrimental project outcomes [Keil et al., 2002; Martin & Chan, 1996; Yetton et al., 2000]. Poor planning can result in unrealistic deadlines or budgets, or poorly defined project goals and objectives.

Once systems development is underway, project management invariably involves managing and controlling resources in the pursuit of project objectives. Time and cost targets may be adjusted to reflect changes in both the project and the organizational context in which it is taking place [Clegg et al., 1997]. Monitoring and control, providing feedback to the project team (e.g. through regular project review meetings), providing them with adequate information and the opportunity to make suggestions relevant to the project (e.g. on project goals and objectives, status, any changes, user needs), coordination of multidisciplinary project teams, and coordination and collaboration with organizational units or groups affected by the system, are all perceived to be important factors influencing system success [Barki et al., 2001; Butler & Fitzgerald, 1999b; Jiang et al., 1996; Jonasson, 2002; Kim & Peterson, 2003; Pan et al., 2004; Peterson et al., 2002; Schmidt et al., 2001; Wang et al., 2006]. In a longitudinal case study, Butler & Fitzgerald [1999b; 2001] found that developers and user representatives felt that regular project meetings enabled project members to keep abreast of each others' activities and of external issues, and that they were good for morale. User representatives felt that the project meetings also enabled them to feel part of the team.

Use of a formal project management method, project management techniques or quality control standards are believed to facilitate the project management process [Barki et al., 2001; Johnson et al., 2001; Kautz et al., 2004]. Based on a survey of systems project managers, Gowan & Mathieu [2005] found that enterprise-wide system upgrade projects are more likely to be completed by their target completion date when a formal project management method is used. They also found that there was a greater need for project management interventions in larger or more technically complex projects. However, Clegg et al. [1997] caution that project management methods and techniques are often criticized for their techno-centric and bureaucratic effects and their neglect of human and organizational issues.

A number of studies have emphasized the importance of having an experienced and competent project manager or leader with both strong technical and interpersonal skills [Coughlan et al., 2003; Jiang et al., 1996; Jiang, Klein et al., 2002; Johnson et al., 2001; Kappelman et al., 2006; Keil et al., 2002; Kim & Peterson, 2003; Peterson et al., 2002; Schmidt et al., 2001; Sumner et al., 2006; Verner & Evanco, 2005; Wang et al., 2005; Wastell & Newman, 1996]. Project leaders can play an influential role in shaping working conditions through their decision making and their ability to motivate and empower the project team [Jiang, Klein et al., 2000; Linberg, 1999; Sumner et al., 2006; Verner & Evanco, 2005; Wang et al., 2005]. However, it has also been suggested that the project manager or leader needs to be able to balance his or her controlling activities with recognition of the autonomous self-control of the project team [Kim & Peterson, 2003; Vidgen et al., 2004].

Several authors emphasize the role of the project manager in mediating between the various groups involved in the project. This might include communicating and translating business and technical requirements between different disciplines [Coughlan et al., 2003; Johnson et al., 2001; Standish Group International, 2001], building consensus and commitment among the project stakeholders [Jiang, Klein et al., 2000; Pan et al., 2004], or acting as a buffer between the project team and external influences [Linberg, 1999]. Verner & Evanco [2005] found that changing the project manager during a software development project was significantly negatively correlated with project success.

**6.3 Use of a Standard Method**

A standard method of software systems development is a formal or documented set of procedures for directing or guiding development, whether commercially or publicly available, or developed internally by an organization. The use of "method" as referring to the codified systematic conduct of systems development is primarily European. North American usage tends to refer to a method as a "methodology" [Iivari et al., 2000/2001; Robey et al., 2001]. Each standard method embodies a set of guiding principles and is based upon a particular philosophy, paradigm or approach to systems development. Usually, each method is supported by a set of preferred development techniques and tools [Fitzgerald et al., 2002; Iivari et al., 2000/2001; Iivari & Maansaari, 1998; Robey et al., 2001; Wynekoop & Russo, 1997].

According to much (although not all) of the literature, use of an appropriate standard method of systems development can improve both the development process and its outcomes, particularly in large or complex projects [Butler & Fitzgerald, 1999b; Fitzgerald, 1998c; Kim & Peterson, 2003; Peterson et al., 2002]. A standard method can facilitate the development process by providing an element of control over aspects such as the sequence of development activities, project management, cost allocation, project team composition and user participation [Lyytinen & Hirschheim, 1987]. Lack of or inappropriate use of a standard method has been considered to increase the risk of project failure [Pan et al., 2004; Schmidt et al., 2001]. A number of studies, however, have failed to find a significant association between the use of standard methods and project success [Barry & Lang, 2003; Fitzgerald, 1998a; Sawyer & Guinan, 1998]. Relative to other factors influencing systems development, use of a standard method has not usually been regarded as a primary mechanism for improving project outcomes, and may not be enough in itself to ensure success of a project [Barry & Lang, 2003; Warne & Hart, 1996]. Kiely & Fitzgerald [2003] suggest that while standard methods do not solve all systems development problems, they can be of help if used properly by experienced developers.

In a survey of development and deployment professionals in the United Kingdom examining the economic impact of using methods on systems development, Chatzoglou [1997] found that using any method was generally better than using no method at all. For example, the use of a method for the entire development process reduced the elapsed time, effort and cost of development, and slightly reduced the number of people involved in development. The use of a method required fewer iterations of the requirements analysis process. Because different methods had different economic impacts, Chatzoglou [1997] suggests that specific methods can help to achieve the desired economic result.

Even where a method is used, the benefits to be had may depend on the context in which it is used [Fitzgerald, 1998b]. In some situations, such as small organizations or small projects with a small development team, the use of methods may hinder rather than help development [Kiely & Fitzgerald, 2002, 2003]. Different groups may also have different perceptions of the relative value of using a standard method. In a Delphi study of software project managers and users, Keil et al. [2002] found that system users perceived that the lack of an effective development process or method was the most important risk to a software systems project (as it would result in a product that did not meet their needs), whereas it was not perceived to be a risk by project managers. They note that users appear to be concerned that project managers may not use an effective development method, while project managers are more confident in their chosen method.

According to Robey et al. [2001], standard methods can influence social interactions and relationships between project participants, by, for example, assigning roles and responsibilities, and indicating how such roles are to interact. Within systems development, participants who control development are in a position to allocate resources and make decisions that may further their own interests. Certain methods place developers in control of development, while others place users or other stakeholders in control. Robey et al. [2001] go on to suggest that the effects of standard methods on social behaviors among participants can potentially influence the outcome of a project. For example, the use of certain systems development tools and methods can improve the work and (therefore) the status of development professionals, but not necessarily other groups, who may consequently feel that the system does not address their work needs as effectively as it might [Lyytinen & Hirschheim, 1987].

## 6.4 User Participation

The term 'user participation', as distinct from user involvement [Barki & Hartwick, 1994], is commonly used as a generic label describing the activities performed by users or their representatives in systems development. A number of authors have reviewed or conducted meta-analyses of prior empirical studies examining the relationship between user participation and system success [Hwang & Thorn, 1999; Mahmood et al., 2000]. Overall, it seems that while in some studies user participation was found to positively influence system outcomes, many studies were inconclusive regarding this issue.

In terms of specific empirical studies, a number have identified a significant positive relationship between user participation and system success [Coombs et al., 1999; Doherty et al., 2003], user satisfaction or acceptance [Foster & Franz, 1999; Hardgrave et al., 1999; Lin & Shao, 2000; Lu & Wang, 1997; Terry & Standing, 2004], project completion [Wixom & Watson, 2001; Yetton et al., 2000], project performance [Aladwani, 2000; Procaccino et al., 2005], system impact [Lynch & Gregor, 2004] or data quality [Zeffane & Cheek, 1998]. Case study evidence also suggests that active user participation is an important component of successful systems development [Butler, 2003; Kim & Pan, 2006; Sumner, 2000; Wastell & Newman, 1996; Wilson et al., 1997].

Perhaps more importantly, various groups of organizational participants perceive user participation to be important to system success, including development managers, systems developers, users, and user managers [Butler & Fitzgerald, 1999b; Fitzgerald, 1998a; Johnson et al., 2001; Kim & Peterson, 2003; Lemon et al., 2002; Peterson et al., 2002; Standish Group International, 1999, 2001]. Similarly, lack of user participation is perceived to be a project risk factor, contributing to system failure or abandonment [Clegg et al., 1997; Johnson et al., 2001; Keil et al., 1998; Keil et al., 2002; Pan, 2005; Peterson & Kim, 2003; Schmidt et al., 2001]. Howcroft & Wilson [2003] describe an organization in which user participation became so entrenched in the systems development culture that it was inconceivable that a software system project would be developed without the participation of users.

At what stages in the systems development process user participation occurs also impacts on project outcomes [Lin & Shao, 2000; McKeen & Guimaraes, 1997; Saleem, 1996]. Empirical studies have shown that user participation in the early stages of development can have greater impact on user acceptance of a system than participation at later stages [Foster & Franz, 1999; Kujala, 2003; Pan, 2005]. Participation throughout the entire systems development process may also encourage user acceptance of the system [Butler & Fitzgerald, 1999b, 2001].

Nevertheless, there have been projects undertaken without user participation that have succeeded and other projects incorporating participation that have not been successful. For example, Gallivan & Keil [2003] present a case study of user participation in a project and its subsequent redeployment and redesign, spanning some twelve years. Despite high levels of user participation throughout this time, users' perceptions of the usefulness of the software system and their actual usage of it remained low, and the project was eventually terminated.

User participation is believed to play a role in shaping users' perceptions of the system in a number of ways. By participating, users can gain an understanding of the system being developed, which may influence their expectations about how the system will serve their interests. This may ultimately avoid a mismatch between user expectations and the delivered system. User participation is also believed to improve users' attitude to and involvement with the system, although the relationship may be influenced by the nature and degree of participation (e.g. voluntary or mandatory participation, direct or indirect participation, and the level of actual influence on design decisions). In other words, it is important that participation is meaningful, involving significant consultation, communication, personal autonomy, decision-making, responsibility and control [Butler & Fitzgerald, 1997; Gallivan & Keil, 2003; Hunton & Beeler, 1997; Kirsch & Beath, 1996; Lynch & Gregor, 2004; Saleem, 1996; Wilson, 2002; Wilson & Howcroft, 2002; Wilson et al., 1997].

By playing an active role in development, users form a sense of ownership of the project and are more likely accept the developed system [Irani et al., 2001; Keil et al., 1998; Myers & Young, 1997]. To this end, users outside the group of user representatives may also need to feel involved and that their interests are being adequately conveyed by their representatives [Butler & Fitzgerald, 1997]. Participation may cause users to become attached to (involved with) the system solution developed, even though it may not meet other criteria for outcome success, such as meeting organizational needs or the needs of other users [Myers & Young, 1997]. However, even with user participation, user resistance may still occur [Butler & Fitzgerald, 2001].

Another important aspect of user participation is that it provides a forum for interaction and communication between users and other groups, in particular systems developers. Such interaction is perceived to enable users to articulate their interests, objectives, and needs; to facilitate the mutual exchange of views and expectations, improving user-developer understanding; and to assist in constructive conflict resolution [Markus & Mao, 2004].

A number of authors have emphasized the importance to system success of active participation in the systems development process of wider groups with an interest in the system (including groups external to the organization), particularly in modern software systems development contexts [Chang, 2006; Jiang, Chen et al., 2002; Liebowitz, 1999; Markus & Mao, 2004; Newman & Sabherwal, 1996; Pan, 2005; Pan & Flynn, 2003; Ravichandran & Rai, 2000; Roberts et al., 2000]. In this way, the interests and objectives of each group may be represented or articulated, mutual understanding may be facilitated, any issues or concerns that arise may be addressed, and commitment (particularly from senior management) may be maintained for the duration of the project [Jiang, Chen et al., 2002; Newman & Sabherwal, 1996; Pan, 2005; Ravichandran & Rai, 2000]. Participation of groups external to the organization (e.g. vendors or external consultants) may provide access to knowledge (e.g. about emergent technologies) that may not be available within the organization [Ravichandran & Rai, 2000]. Failing to include all interested groups, including non-represented user groups, in the system's development can result in a system that does not address their needs, or can lead to their lack of commitment or active resistance to the system [Pan, 2005; Pan et al., 2004]. For example, Pan [2005] describes the development of an electronic procurement system in which the procurement manager ignored the concerns of the organization's suppliers who felt their business interests were threatened by the new system. The perceived threat of the new system united the suppliers in influencing the organization to abandon the new system.

## 6.5 User Training

User training and education has been identified in the literature as a further factor that can influence the outcome of a software systems development and deployment project. A number of studies have found that user training can be important for system success [Coombs et al., 1999; Riley & Smith, 1997; Skok & Legge, 2002; Sumner, 2000; Wastell & Newman, 1996], although it may be time-consuming in some large projects [Mabert et al., 2003].

Training seems to affect project outcomes through its influence on users' attitudes towards the system. Through a training program, users can gain skills and experience in utilizing the system, potentially increasing their confidence in using it, as well as greater knowledge and understanding of the system, which can influence their acceptance (or rejection) of it [Skok & Legge, 2002]. It has been argued that user education and training may be critical to the long term success of a system, especially when users feel threatened (such as by changed job roles), as incomplete knowledge and understanding of the system and a lack of appreciation of changes can lead to resistance towards new systems [Irani et al., 2001; Marion & Marion, 1998]. Wilson & Howcroft [2002] argue that training can also be used to try to persuade users of the benefits of a new system in an effort to enroll them to use it. Although training usually begins after installation has occurred [Jiang et al., 1998a], Mahmood et al. [2000] suggest that by introducing a training program earlier in the development process users may contribute more effectively to development.

## 6.6 Management of Change

The management of changes resulting from systems implementation has long been recognized as important to the outcome of software systems projects [Lyytinen & Hirschheim, 1987]. The introduction of a system to an organization can produce considerable changes and have consequences for many users of the new system [Butler & Fitzgerald, 1997; Riley & Smith, 1997]. Confronted with change, individuals may experience a range of negative emotions such as fear, anger or denial. They may be reluctant to share their knowledge or information, or may provide inaccurate or conflicting information, if they feel that their jobs are threatened. They may resist changing how they work or even resist using the new system [Butler, 2003; Butler & Fitzgerald, 1999b, 2001; Coughlan et al., 2003; Lin & Shao, 2000; Lu & Wang, 1997; Olesen & Myers, 1999; Pan, 2005; Pan et al., 2004; Skok & Legge, 2002; Wixom & Watson, 2001]. According to van Offenbeek & Koopman [1996], potential resistance increases when the individuals involved have a low potential for change, a low willingness to change, and when the impact on the organization is high.

Risk management notwithstanding, some consequences cannot be anticipated or identified at the start of a project. Increasingly sophisticated, flexible and integrated systems increase the potential for unpredictable or unintended consequences [Doherty et al., 2003; Robey & Boudreau, 1999]. Further, individuals may interpret or appropriate the system in a variety of ways during its development and use [Eason, 2001].

While change management is not necessarily an issue in every project, many studies highlight the ongoing importance for system success of addressing organizational change, or the perils of ignoring or inadequately understanding the dynamics of change that occur for both individuals and the organization [Butler, 2003; Butler & Fitzgerald, 1997, 1999b, 2001; Dhillon, 2004; Irani et al., 2001; Kappelman et al., 2006; Lu & Wang, 1997; Schmidt et al., 2001]. Systems development can overlook organizational changes, such as changes to structures and processes, workloads, organizational roles, job content or autonomy [Clegg et al., 1997; Doherty et al., 2003]. Dhillon [2004] argues that a consideration of power relationships within an organization during systems design and implementation is essential in order to manage the alignment of these consequential changes.

Several authors suggest that change management issues need to be addressed and resolved early in the systems development process to avoid subsequent problems [Butler & Fitzgerald, 1999b; Skok & Legge, 2002]. Eason [2001] notes that even when change management practices are well established, they tend to occur after systems design, restricting the possibilities for social or organizational issues to be taken into account. With respect to enterprise-wide systems, Skok & Legge [2002] recommend that organizations need to act to change the culture within the organization, possibly starting long before the new system is implemented. Enterprise-wide systems can involve significant changes, such as changing business processes, organizational structure and culture; altering data ownership, use and access; or changing roles, work processes and jobs specifications [Chang, 2006; Doherty & King, 1998b; Doherty et al., 2003; Irani et al., 2001; Riley & Smith, 1997; Skok & Legge, 2002; Wixom & Watson, 2001].

As noted earlier, a number of authors suggest that managers within an organization, particularly top management, can play an important role in facilitating system-related change by championing the project, creating a suitable context for change, and countering any negative attitudes [Butler & Fitzgerald, 1999b; Kim & Peterson, 2003; Lemon et al., 2002; Riley & Smith, 1997; Wixom & Watson, 2001; Yetton et al., 2000]. However, some managers may be reluctant to challenge what they perceive as powerful user groups [Doolin, 2004; Marion & Marion, 1998; Riley & Smith, 1997; Wilson, 2002]. Development professionals may also play an important bridging role in managing change by facilitating communication between different participants in a project [Marion & Marion, 1998]. Symon [1998, p. 39] emphasizes the role of internal systems developers as change agents, "effectively embedD new organizational systems into organizational contexts".

Key properties or attributes of the influences that comprise the development processes dimension are summarized in Figure 5.

**Figure 5:** Properties/attributes of influences related to development processes

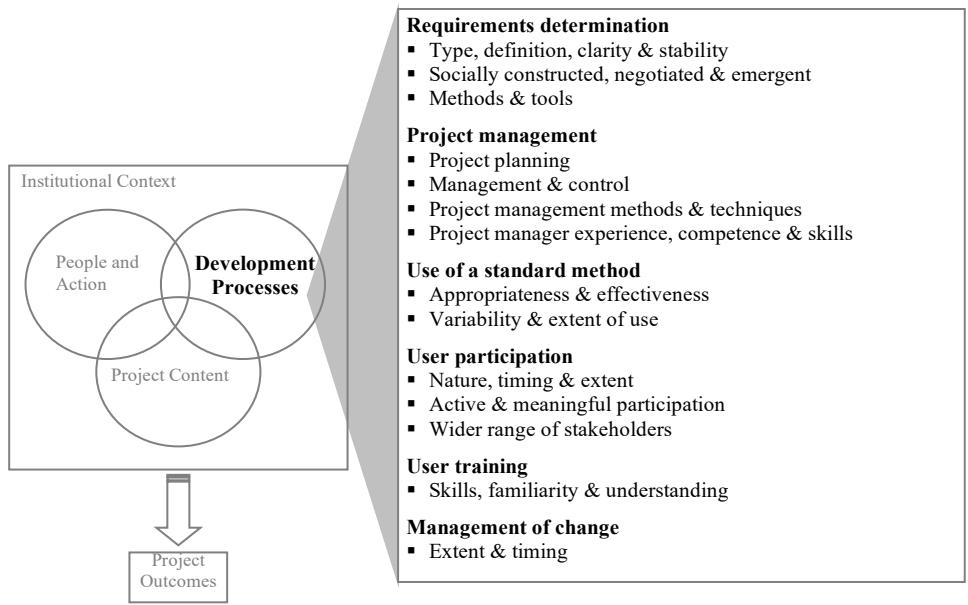

## 7. INFLUENTIAL FACTORS – INSTITUTIONAL CONTEXT

Various authors have argued that contextual properties of the organization, as well as the wider social, economic, political, cultural, and historical environmental conditions in which an organization is located, can influence systems development and deployment project outcomes, often in unpredictable ways [Bussen & Myers, 1997; Constantinides & Barrett, 2006; Gärtner & Wagner, 1996; Iivari, 2004a; Mitev, 2000]. Unlike many of the other factors discussed above, these contextual factors often lie outside the direct control of the project team [Bussen & Myers, 1997]. Systems development occurs across layers of context, ranging from the local organizational context to the national and international environment [Avgerou, 2001; Christiaanse & Huigen, 1997; Krishna & Walsham, 2005; Symon, 1998]. Elements of the institutional context, comprising both internal organizational properties and external environmental conditions, may shape project outcomes through their influence on systems development processes and procedures, such as user participation and standard method use. These elements may include an organization's structures, culture and practices, the historical context of system use within the organization, and wider socio-economic conditions and regulatory requirements.

### 7.1 Organizational Properties

The term 'organizational properties' is used here to encompass a range of organizational structures, practices and relations that make software systems development possible. These include: institutionalized norms, values and beliefs; the distribution of available organizational resources (time, money and skills); standard rules and operational procedures; established customs and practices; formal and informal organizational structures; control and coordination mechanisms; reward structures; and the division of labor [Knights & Murray, 1994].

Particular structural properties and context-specific features can enable or constrain the course of systems development. For example, structures of authority within an organization will influence the time, money, tools and other resources available for development, such as organizationally-imposed restrictions on system-related expenditure [Bussen & Myers, 1997]. Organizational structure and culture may discourage or encourage communication and cooperation between functional units [Gallivan & Keil, 2003]. According to Butler & Fitzgerald [2001], the increased size and complexity of mature software systems development groups in older or larger organizations may decrease their ability to develop systems that are perceived as useful.

Three types of organizational properties that have received particular attention in the literature include aspects of organizational culture, those related to organizational policy and established practices, and the history of systems development and use in the organization.

### 7.1.1 Organizational culture

Organizational culture can be viewed as a symbolic system of learned and shared sets of meanings that provide patterns for behavior within an organizational setting [Iivari, 2004a; Walsham, 1993]. Relevant aspects of organizational culture include systems of ideas and symbols, values and beliefs, collective identity, shared experiences, and common understandings, interpretations and assumptions that shape behavior or action in relation to systems development and implementation [Iivari, 2004a; Robey & Boudreau, 1999].

The established organizational culture may reflect widely accepted norms and values that influence interactions between users and developers, inter-departmental cooperation, or the intended use of a software system [Nandhakumar & Avison, 1999; Nandhakumar & Jones, 1997; Olesen & Myers, 1999; Somers & Nelson, 2001]. For example, an organizational culture based on consensus encourages communication and conflict resolution [Coughlan et al., 2003], facilitating positive project outcomes. Umble et al. [2003] highlight how the development of an organizational culture that was receptive to change and continuous improvement facilitated implementation and acceptance of the changes associated with an ERP system. In a contrasting example, Olesen & Myers [1999] describe how the existing culture and norms of an organization meant that users appropriated a new groupware system in a way that reproduced their existing work practices rather than accepting the work-related changes envisaged by senior management.

In her study of three software development organizations, Iivari [2004b] identified multiple discourses on user participation that constructed user participation in different ways in the organizations. Butler [2003] describes the systems development practices in a large multinational organization where the social matrix and identity of the organization (including culture, structure, business processes, communication and learning) were shaped by the dominant group of employees who were engineers. Engineering 'communities of practice' existed within the various business functions and retained a high degree of autonomy in developing their own systems.

In another example of the influence of organizational culture on systems development, Chae & Poole [2005] discuss the development of an enterprise-wide system in a university context. The new system was envisaged by the project sponsors and project team as a centralized, integrating system that could serve as a standard solution across different sized organizations and various levels of users. Centralization would mean that it would be easier to modify the system in response to external changes, such as new regulations or laws. However, the system was developed locally within particular units rather than globally, in an organizational culture that emphasized decentralized decision making and autonomy. As development proceeded, the project team modified their development approach to become more user-oriented in an attempt to satisfy the unique needs of the various organizational units. According to Chae & Poole [2005], the result was an 'average' system that satisfied nobody. Some units customized the new system using workarounds; other units continued using existing systems or developed yet other alternatives.

In studying the development of an EIS, Nandhakumar & Jones [1997] found that while established hierarchical organizational structures initially restricted opportunities for interaction between developers and executive users, they also provided a medium for some legitimate interaction. In conforming to such established patterns, individuals reproduce the norms and values that underlie them. However, individuals may also be able to modify established patterns of behavior, or at least find ways of working around those that are relatively resistant to change (e.g. using intermediaries such as secretaries as a source of user requirements) [Nandhakumar & Jones, 1997].

### 7.1.2 Organizational policy and practice

There is a link between organizational culture and the policies and practices which emerge around systems development. Robey & Newman [1996] suggest that organizations may have an embedded cultural orientation to systems development (or even sub-cultures with different perspectives). They argue that "cultures develop rituals that are repeated, and systems development can be regarded as a ritualistic cultural practice" [p. 59]. An organization's policies or procedures can enable or constrain individuals' actions by enforcing organizational rules or norms of what constitutes appropriate or acceptable behavior [Butler, 2003]. In this way, organizational policies and established practice related to software systems development may define and thereby influence human action in development activities [Butler & Fitzgerald, 2001]. Once a particular practice has been utilized on a routine basis, it becomes institutionalized (or taken-for-granted), becoming an integral part of the organization's culture [Butler & Fitzgerald, 1997].

Charette [2005] suggests that increasing numbers of organizations are assessing their development practices using approaches such as the Capability Maturity Model (CMM), and its variants, for development, acquisition and for people. Such an approach reflects an organizational culture that first seeks to have defined and repeatable processes before possibly building on these towards continuous improvement and optimization. The motivation for organizations to adopt such an approach may draw on several factors; e.g. these may be financial (in terms of being in a position to be awarded contracts), or cultural (in terms of embracing the principles of continuous improvement). Interestingly, there are to date very few empirical studies that have identified process maturity as an influential factor in affecting project outcomes, at least in terms of the search undertaken here.

Existing organizational policies and practice may constrain the appropriation of systems development innovations, such as new standard methods, techniques or tools. In a case study of system design, Gasson [1999] found that even though attempts were made to utilize a new approach to design (integrating business process investigation with technical system design), established practice continued to influence systems development. It did so by constraining the choices of available methods and tools, and informing the problem-solving approach of the 'expert' designer on the project team, who initially tried to impose a structured approach on systems development.

However, organizational policies on and practice in systems development can change over time [Heiskanen et al., 2000; Robey & Newman, 1996]. For example, drawing on two projects in a large Irish organization, Butler & Fitzgerald [2001] illustrate how the organization's policy on user participation and development-related change influenced how user participation and change management were enacted. The organization had a participative approach to decision-making and change, which was reflected in their policy and institutionalized practice of user participation. Both projects had high levels of user participation but still experienced change-related problems. As a result of the problems experienced, the organization implemented a more structured policy on development-related change and negotiated employee commitment to future changes. The organization's policies and procedures in relation to systems development continued to evolve, in response to either past experiences or to changes in the systems development context [Butler & Fitzgerald, 1999a].

### 7.1.3 Organizational systems history

Knights & Murray [1994] discuss various aspects of technology which form local conditions that may influence systems development in an organization. These include attitudes to and understandings of systems within the organization; the position occupied by development specialists within the organizational structure; and the legacy and past experience of systems development and use. For example, a history of system failures in an organizational context can create cynicism or resistance towards new systems development [Doolin, 2004]. On the other hand, success in prior projects within an organization does not necessarily guarantee success in future projects [Goldstein, 2005].

Various authors have suggested that the analysis of system failures (and by analogy, system successes) by organizations can potentially play an important role in informing software systems development practice (e.g. by supporting established practice or suggesting changes) [Lyytinen & Robey, 1999; Nelson, 2005; Poulymenakou & Holmes, 1996; Warne & Hart, 1996]. However, Lyytinen & Robey [1999] argue that many organizations fail to learn from their previous systems development experiences. By ignoring or reinterpreting relevant information, they have learnt to fail to the point that failure comes to be accepted as normal. If this situation continues, failure itself can become institutionalized. For example, in a case study describing the development of an electronic procurement system at a local government organization in the UK, Pan et al. [2004] found that failure had become an acceptable norm.

An organization's failure to learn may arise from limited time available for reflective analysis, a reluctance to allocate additional resources for retrospective analysis of a failed project, a desire to move on, a high turnover of staff with relevant experience and knowledge, and established institutionalized arrangements and patterns of thinking. There may be no incentives to learn from system failures; in fact, organizations may try to forget their failures or punish those perceived to be responsible for them. Further, organizational structure or competition between business units may inhibit interaction, information sharing and learning between groups involved in a project failure [Lyytinen & Robey, 1999; Nelson, 2005]. Lyytinen & Robey [1999] discuss three generic 'myths' that inhibit learning from failure. These are the myth of the 'technological fix', in which more and better technology will solve systems development problems, the 'organizational' myth that changing organizational design (e.g. changing the organizational structure, outsourcing or process re-engineering) will overcome development challenges, and the 'silver bullet' myth, in which a 'magical' solution exists that will rectify difficulties encountered in systems development.

Legacy systems and an organization's existing technological infrastructure can also influence the development of software systems [Knights & Murray, 1994]. According to Chae & Poole [2005], pre-existing systems (both internal and external to an organization) play an active role in shaping the direction of new systems development. Drawing on a case study of the development of an enterprise-wide system, they argue that pre-existing systems can exert an influence by constraining or directing the new development trajectory. For example, the new system described by Chae & Poole [2005] had to conform to the requirements of the existing computing infrastructure in the organization and other systems with which it was meant to interface and exchange data. In considering design options, the project team took account of alternative systems in other organizational settings, which acted as standards of functionality for the new system. Pre-existing systems can also affect approaches to developing a new system through developers' prior experiences and learning. For example, Chae & Poole [2005] describe how the project team director adopted a relatively conservative approach to the project that was influenced by his previous experiences in developing large-scale systems. Similarly, Symon & Clegg [2005] observe that the history of systems development in an organization can influence the strategy adopted for user participation in software systems projects.

## 7.2 Environmental Conditions

Knights & Murray [1994] discuss the general and local socio-political and economic conditions within which an organization functions. They suggest that "within a market economy, these conditions largely concern labor, product and capital markets, their respective regulatory frameworks, and the social relations of class, gender and race" [p. 43]. Bussen & Myers [1997] describe the case of failure of an EIS in a large organization. While their case exhibited many of the traditional risk factors identified within the academic literature, the authors also identify various environmental conditions, which they argue probably had more influence over the project outcome. These included changes in company ownership, leading to eventual overseas ownership, and rapid organizational and economic growth in a depressed economy. Changes in the external environment may also mean that a proposed system loses its former relevance [Doolin, 2004].

A range of external entities operating at the environmental level can influence systems development decisions and practices. These potentially include: government authorities, international agencies, professional and industry associations, trend-setting and/or multinational corporations, universities, financial institutions, and trade unions. For example, the impetus to introduce a new system may arise from a new government initiative [Doolin, 2004; Myers & Young, 1997]. In fact, a new system may be the means by which the policies or objectives of government are imposed on an organization [Myers & Young, 1997]. Conversely, withdrawal of government financial support may result in project abandonment [Constantinides & Barrett, 2006; Doolin, 1999].

External entities exert their influence through a range of processes such as building and/or deploying specific knowledge related to systems development; subsidizing or directing development; establishing standards, norms or regulations within which systems development occurs; and institutional isomorphism [Avgerou, 2001; Nicolaou, 1999]. Institutional isomorphism, the idea that organizations in the same field adopt similar structures and processes, may occur through coercive pressures, such as government mandates, industry standards or dominant business partner influences [Chae & Poole, 2005].

Isomorphic effects can also be seen in the voluntary imitation of organizations' system development processes and decisions that are perceived to be successful, or in the normative effects of professional networks and educational institutions [Nicolaou, 1999]. For example, system developers work within professional disciplines, which represent "bodies of knowledge that preserve concepts, practices, and values" [Chae & Poole, 2005, p. 23]. These disciplines structure developers' actions in the systems development process.

Differences in national cultural contexts may cause a range of issues in systems development, including attitudes to project roles, use of procedures, developer autonomy, team relationships, flexibility for organizational or process change, and the balance between technical and organizational issues [Coughlan et al., 2003; Krishna & Walsham, 2005]. Walsham [2002] explores contradiction and conflict in a case study of cross-cultural systems development work. He suggests that the conflict that developed around management style, work ethos and project coordination reflected "differences in deep-seated cultural attitudes" [p. 365]. Similarly, Kumar et al. [1998] discuss how traditional US systems development approaches based on technical-economic rationality do not translate well into different cultural contexts, which may require consideration of specific cultural dimensions of work and communication practices.

The influence of national culture can also be seen in Mitev's [2000] description of the difficulties encountered by the French government rail service in introducing a computerized reservation system originally developed for the US airline industry. The new system completely changed the established practices of rail workers and railway users, who rejected such changes. According to Mitev [2000, p. 90], the difficulties arose through attempts to translate "management discourses, commercial practices, economic models, strategic goals, political perspectives, sectorial markets, and structures" between two very different cultural and sectorial contexts.

Key properties or attributes of the influences that comprise the institutional context dimension are summarized in Figure 6.

**Figure 6:** Properties/attributes of influences related to the institutional context

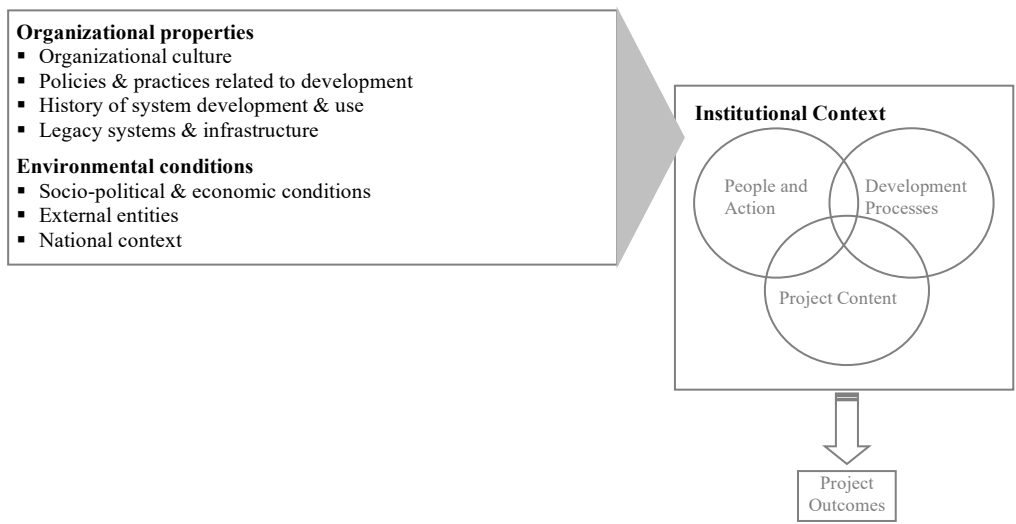

# 8. DISCUSSION

From the preceding survey of the research literature, five general themes emerge. These relate to the persistence of certain traditional factors influencing software systems development, the influence of the changing nature of development, the relative importance of people and process in project outcomes, the recognition of the importance of the institutional context in which software systems development takes place, and the need to focus on the interrelationships and interactions between factors influencing software systems projects.

## 8.1 The More Things Change, the More They Stay the Same

A number of factors highlighted in literature prior to the review period as affecting software systems project outcomes (e.g. see Lyytinen & Hirschheim [1987]) continue to be perceived, and empirically demonstrated, as important positive influences on contemporary systems development. These are probably best regarded as necessary but not sufficient for achieving positive project outcomes. They include:

*People and action*

- developers with adequate experience, application domain knowledge and interpersonal skills;
- committed users with realistic expectations of the system;
- committed and supportive top management;
- effective functioning of the project team;

*Project content*

- clear, well-defined and well communicated project goals and objectives;
- adequate time, financial and human resources;
- the use of appropriate technology;

*Development processes*

- well defined and clearly stated user requirements;
- the use of an appropriate standard method of systems development.
- the active participation of users in systems development; and
- adequate user training.

Many of these factors have become well established in the software systems development culture, and are frequently rehearsed in the software systems practitioner literature [e.g. Charette, 2005; Jurison, 1999; Reel, 1999]. What is difficult to explain is why, despite the apparent knowledge of these factors in software systems development practice, does software system project failure continue to occur? As Cobb's Paradox states, "We know why projects fail, we know how to prevent their failure – so why do they still fail?" [Royal Academy of Engineering, 2004, p. 10].[2]

While it is tempting to place responsibility for this situation on a failure to adhere to best practices, there may be other possible explanations. Sauer [1999, pp. 291-292] provides a useful analysis of why organizations apparently "continue to do the things identified as factors associated with or causing failure". He suggests that either the 'true' causes of software systems project failures have not yet been identified or, more likely, the various factors are causes of failure but are not readily avoidable. In either case, he criticises prescriptive, factor-based research on project failure for four reasons.

First, Sauer [1999] suggests that most prescriptions lack specificity. For example, the 'adequacy' of resources and training, the 'appropriateness' of development technology and methods, or the 'clarity' of goals and system requirements, typically remain undefined in prescriptive lists of project 'success' factors. Critically, such evaluations are only made post hoc and, in a circular argument, in reference to the perceived success or failure of the project [Sauer, 1999]. Similarly, while 'user participation' has become a routine prescription for systems development, exactly who is a 'user' and what actually comprises 'active' user participation is often not specified. Second, Sauer [1999] suggests that some prescriptions are not easily acted upon. He uses the example of the importance of top management support, pointing out that its absence is difficult to measure and that gaining it is often difficult to achieve in practice. Third, Sauer [1999] suggests that organizational or environmental conditions may inhibit whether a prescription can be followed in practice. Finally, he suggests that prescriptive 'cures' may exacerbate other problems in software systems projects. For example, an unqualified prescription for top management support may lead to escalation of commitment to a failing course of action [Keil & Robey, 2001].

Finally, prescriptive lists of generic factors also imply that each factor is independent, universally applicable, and of equal importance in specific software systems projects. In practice, the influence of factors is temporal in nature. Rather than being "frozen in time" [Nandhakumar, 1996, p.62], factors may vary dynamically in their relative importance and influence at different times during the course of a project. This suggests that different factors may be significant, and thereby require explicit attention, at particular times or stages [Nandhakumar, 1996; Somers & Nelson, 2001, 2004]. In addition, several authors have conceptualized factors as operating from within different layers of a multilayered context, suggesting that factors from different layers will vary in the magnitude and frequency of their impact [Nandhakumar, 1996; Scott & Vessey, 2002]. Moreover, it is likely that factors in a particular project context involve complex interrelationships and interactions. As Sauer [1999] observes, this complexity makes theorizing about software systems project outcomes difficult.

---

[2] We are grateful to an anonymous reviewer for bringing Cobb's Paradox to our attention and for highlighting the potential for interactions among factors.

The continued emphasis given to the factors listed above in the systems literature over a long period of time suggests that they constitute a set of fundamental (but not exclusive) issues that need to be addressed in most projects. However, changes to the nature and practice of software systems development in relatively recent times have brought other issues and factors to the fore.

## 8.2 The Changing Nature of Software Systems Development

Various authors have argued that the nature of software systems development has changed significantly in recent years [e.g. Kiely & Fitzgerald, 2003; Markus & Mao, 2004]. These changes tend to reflect rapid advances or changes in technology, the demands of an increasingly complex, global business environment, and changing systems development practices. In many cases, these changes are inter-related. For example, systems based around new technologies, such as the Web or rich media, have typically involved more flexible, non-traditional development approaches, often *ad hoc* or informal in nature [Avison & Fitzgerald, 2003; Barry & Lang, 2003; Britton et al., 1997; Taylor et al., 2002], although Lang & Fitzgerald [2005; 2006] suggest that Web and hypermedia systems development is more disciplined than previously thought. For example, Bahli & Tullio [2003] discuss the emergence of 'web engineering' – new methods and tools for Web-based systems development projects. Further, differences between traditional and Web-based development projects are likely to become less pronounced over time as the latter are increasingly integrated with other organizational systems [Vidgen, 2002].

Modern software systems development is generally characterized by increasing devolution of development expenditure to business units or user groups, high levels of packaged software acquisition and customization, increased outsourcing of systems development, and concomitant reduced levels of in-house systems development [Avison & Fitzgerald, 2003; Clegg et al., 1997; Fitzgerald, 2000; Keil & Tiwana, 2006; Sawyer, 2001b; Schmidt et al., 2001]. The increase in packaged software acquisition and implementation by organizations, in effect consuming software rather than developing it, has led to changed or new influential factors in systems deployment. For example, increased emphasis is placed on vendor selection and relationships, product feature analysis and comparison, system configuration or customization, and necessary changes to business processes [Sawyer, 2001b; Somers & Nelson, 2001; Umble et al., 2003].

Another aspect of the changing nature of systems development seems to be the development of smaller-sized projects or the delivery of larger projects in parts, which can increase the chances of successful project outcomes [Johnson et al., 2001; Software Magazine, 2004]. Smaller-sized projects are partly a result of factors such as standard software infrastructure use [Johnson et al., 2001], incremental development [Avison & Fitzgerald, 2003], and the need for rapid delivery of systems in the short time frames characterizing the modern business environment [Baskerville & Pries-Heje, 2004; Fitzgerald, 2000].

At the same time, the emergence of enterprise-wide systems, inter-organizational systems and globally distributed systems have led to increased complexity in some development and deployment projects [Bahli & Tullio, 2003; Espinosa et al., 2006; Gowan & Mathieu, 2005; Keil & Tiwana, 2006; KPMG, 2005; Parr & Shanks, 2000; Royal Academy of Engineering, 2004; Wixom & Watson, 2001]. Increased complexity and the concomitant organizational changes associated with such systems can adversely affect project outcomes. This has highlighted the ongoing need to effectively manage such changes, particularly from early in the development process [Eason, 2001; Skok & Legge, 2002]. However, as systems become increasingly sophisticated and integrated, the possibility of unpredictable or unintended consequences also increases [Doherty et al., 2003; Robey & Boudreau, 1999].

Modern systems development contexts are also tending to require the active participation of a wider group of stakeholders in a software project [Markus & Mao, 2004]. For example, the trend towards outsourcing of systems development or the increasing prevalence of enterprise-wide systems introduces new participants such as vendors and outsourcing contractors, and the external consultants who play an increasingly active role in mediating between these participants and the system client [Chang, 2006; Howcroft & Light, 2006; Sawyer, 2001b]. Managing or controlling such parties is becoming increasingly important in systems development [Pan et al., 2004; Schmidt et al., 2001]. The range of activities in which they may potentially participate has also diversified, such as in the process reengineering or change management often associated with the development and implementation of enterprise-wide systems [Markus & Mao, 2004].

There is some evidence that improved project management and having more skilled project managers has led to increased project success [Johnson et al., 2001]. Project management may be assuming a more influential role in systems development, particularly in large or complex, enterprise-wide systems where there is likely to be a relatively greater need for project management interventions [Gowan & Mathieu, 2005; Somers & Nelson, 2001]. Indeed, in a report on the challenges of complex software and IT projects, the UK Royal Academy of Engineering and British Computer Society [Royal Academy of Engineering, 2004] emphasized that the importance of project management is still not well understood and is often underestimated.

## 8.3 People and Process

Difficulties associated with software systems development and implementation are often divided into technical issues and organizational or human-related issues. Historically, technical issues dominated accounts of systems development outcomes. However, in the last decade there has been increasing recognition that project failures are rarely caused by technical problems alone [Clegg et al., 1997; Eason, 2001; Flynn & Jazi, 1998;

Luna-Reyes et al., 2005; Markus & Benjamin, 1996; Markus & Mao, 2004; Poulymenakou & Holmes, 1996; Royal Academy of Engineering, 2004]. A number of studies have recognized the importance of organizational, political and human-related issues, often finding that these are more important than technical issues in determining system outcomes [Clegg et al., 1997; Doherty & King, 1998a, 1998b, 2001; Doherty et al., 2003; Drummond, 1996; Irani et al., 2001; Oz & Sosik, 2000; Riley & Smith, 1997]. As the Standish Group [1999, p. 5] note, "What has become clear … is that people and process have a greater effect on project outcome than technology".

Organizational or human-related issues that contribute to system underperformance include: inadequate or misaligned organizational and business strategies to guide development and implementation; inadequate user participation and ownership of the system; insufficient education and training; insufficient organizational resources or support for organizational or human-related issues; lack of attention to organizational structure, processes, culture and professional autonomy; and lack of attention to job and task design, usability, and user working styles and motivations [Clegg et al., 1997; Doherty & King, 1998a, 1998b, 2001; Doherty et al., 2003; Riley & Smith, 1997]. In a software systems project described by Gallivan & Keil [2003], the reasons given by users for not using the system were technically-based. Even when these shortcomings were addressed, so that the users' perceptions of the problems improved, usage of the system still did not improve. Gallivan & Keil [2003] conclude that the underlying reasons for not using the system were related to a perceived incongruence in task-technology fit, and that these reasons had not changed despite the technical redesign that occurred.

Increasingly, systems development professionals are recognizing the importance of organizational issues, although they still tend to address those issues which have a more technical element than those which are less tangible [Doherty & King, 1998a, 1998b, 2001; Doherty et al., 2003]. This is compounded by the techno-centric nature of much software systems development and the use of standard methods, which tend to encourage consideration of organizational implications *after* system implementation [Clegg et al., 1997; Doherty & King, 1998b]. Many organizations appear to be using *ad hoc* interventions to address organizational issues as they occur (often after implementation), rather than formal socio-technical approaches [Doherty et al., 2003; Eason, 2001].

There is also increasing recognition in the literature that software systems development is a process of social interaction, and that the nature and quality of interactions between participants can strongly influence the course and outcome of a project [e.g. Heiskanen et al., 2000; Wang et al., 2006]. Increasing participation of more stakeholder groups suggests that the problems associated with interaction amongst participating groups, such as differences in goals, expectations, and understanding of the system being developed, may be exacerbated. The potential for conflict around systems development increases when the number and diversity of participants increases, such as in systems that require high levels of integration across an organization or involve external parties [e.g. Yetton et al., 2000]. Some authors are now suggesting that conflict, if resolved, can be beneficial if it leads to constructive debate or better decision making [Robey et al., 2001; Sawyer, 2001a].

Similarly, there is increased recognition that the various groups of participants associated with a software systems development and deployment project are not homogeneous. For example, Jiang et al. [1998b] suggest that developers are heterogeneous in their technical, socio-political and user orientations. Equally, in projects that involve the development or implementation of complex, enterprise-wide or inter-organizational systems there may be multiple user groups or functional units with different interests [e.g. Markus & Mao, 2004; Pan, 2005]. The introduction of a new or changed system may be problematic where it challenges professional roles, autonomy and cultures [e.g. Doolin, 2004; Myers & Young, 1997; Wilson, 2002]. A number of authors have suggested that user representatives may not actually represent the full user community or can be 'captured' by the software development team [Butler & Fitzgerald, 1999a; Gallivan & Keil, 2003].

## 8.4 The Importance of Institutional Context

An area of relative neglect in the software systems development literature that is beginning to receive more attention is the importance of the links between systems development people and process and the institutional contexts in which systems development occurs. A number of authors have emphasized a strong interrelationship between context and action, such that the form, nature and conduct of systems development processes need to be viewed as situated within their contextual setting [e.g. Constantinides & Barrett, 2006; Gasson, 1999].

At the level of the organization, the development of software systems occurs within a context of established organizational practices and structures that guide appropriate behavior in organizational activities. Further, systems development involves multiple interested groups and hence potentially traverses different subcultures or communities of practice, each with distinctive shared beliefs, norms, and understandings. The institutionalization of particular systems development policies and practices, which both facilitate and constrain project outcomes, is the result of negotiation or contestation between these different organizational groups. In particular, the history of software systems development and use in an organization may play an active role in shaping the direction of new developments.

Organizations and their software systems development efforts also exist in a wider social, political and economic environment. Various elements of the wider environmental context may shape the course of development in any given software systems project. These include the influence of labor and market conditions, government regulation or intervention, industry or competitive pressures, and specific dimensions of national culture. Empirical research on software systems projects needs to incorporate a consideration of the way in which project outcomes emerge from their historical and organizational context, together with an appreciation of software systems development in the context of a wider set of social, cultural, political and economic conditions.

## 8.5 Interrelationships and Interaction

As noted in Section 8.1 above, many factor-based studies implicitly assume, or give the impression, that individual factors implicated in software systems projects are independent in their operation and effect. In practice, however, this is not the case. A number of empirical studies have emphasized that software system project outcomes typically involve multiple factors that interact in complex ways, either directly or indirectly [e.g. Akkermans & van Helden, 2002; Butler & Fitzgerald, 1999b; Clegg et al., 1997; Nandhakumar, 1996; Scott & Vessey, 2002]. The relationship between factors is dynamic, varying in terms of the direction, strength, and timing of their influence on each other [Butler & Fitzgerald, 1999b; Nandhakumar, 1996]. Understanding these interactions is likely to be critical to gaining greater insights into how and why software systems project outcomes occur [Nandhakumar, 1996]. Although beyond the scope of this paper, detailed analysis of the interrelationships between factors influencing software systems project outcomes is an area that requires further work. Two broad strategies for attempting this were observed in the recent empirical literature.

Some quantitative empirical studies reviewed in this paper hypothesized and found statistical evidence for relationships between specific factors, indirect effects on software systems project outcomes, or the role of mediating factors on relationships. However, where present, such findings were difficult to synthesize given the lack of clarity and inconsistent treatment of various factors or outcome measures across empirical studies. Future research would benefit from a greater degree of consensus or agreement among the software systems research community over the use of common and explicitly defined terminology, together with instruments and scales used to measure specific factors and project outcomes. Past experience has shown that this may be possible. As part of a move to address methodological problems associated with early studies of the participation of users in software systems projects, Barki & Hartwick [1994] defined two distinct constructs, 'user participation' and 'user involvement', where previous studies had used the terms interchangeably. Subsequent empirical studies [e.g. Hunton & Beeler, 1997; Lin & Shao, 2000; McKeen & Guimares, 1997] and meta-analyses [e.g. Hwang & Thorn, 1999] have used Barki & Hartwick's definitions of these constructs (although other studies have tended to perpetuate the confusion).

A second group of empirical studies addressed the complexity of interrelationships between factors by using process research approaches, which attempt to explain how particular project outcomes develop over time as the consequence of a preceding sequence of interrelated and interdependent events and factors in organizational processes [Markus & Robey, 1988]. These studies use longitudinal, qualitative data to identify simultaneous influential relationships among multiple factors over time. Their analyses are often represented using mapping techniques, such as casual loop diagramming [Akkermans & van Helden, 2002], network analysis [Butler & Fitzgerald, 1999] and influence diagrams [Kim & Pan, 2006]. The aim is to illustrate the complex interrelationships and influences between factors, while avoiding the reductionism of other research approaches [Butler & Fitzgerald, 1999]. Kim & Pan [2006, p. 63] suggest that such an approach "facilitates linking pieces into a whole picture, and interpreting the influence of any one factor on others. This in turn facilitates an understanding of the chain of events that link the factors to success".

There is a need for empirical research that undertakes a more in-depth consideration and conceptualization of software systems development. After all, "systems development is, in essence, a multi-dimensional change process that takes place … [within] a complex web of social conditions and factors that shape and influence the … development process and its outcomes" [Butler & Fitzgerald, 1999, pp. 351-352].

## 9. CONCLUSION

We have presented an extensive survey and synthesis of recent literature (1996 to 2006) addressing factors influencing the outcomes of software systems development and deployment projects. The intent of the survey was to consider whether or not these factors had changed relative to traditional studies of systems development, possibly in line with changes in development methods and practices, and to provide the basis for a contemporary and inclusive analytical framework that would facilitate ongoing investigation of influential factors. The framework presented in Figure 1 incorporates four groupings of factors – people and action, project content, development processes, and institutional context – that together enable a project and its outcomes to be considered in terms of content, process, action and context [Walsham, 1993]. The result is an empirically grounded framework that reflects contemporary thinking, recognizing that the development and deployment of software systems is a multi-dimensional process in which people and technology act and interact in locally situated contexts.

The classificatory framework serves as a useful analytical device for categorizing and synthesizing the empirical literature on factors influencing software systems development. It can be used by researchers as a device for enabling analytical abstractions, since the four dimensions of the framework are conceptually more manageable than having to deal with the eighteen individual factors identified by our review. When greater detail is required, consideration can be given to the individual factors within each dimension. A particular strength of the framework is its capacity to accommodate interrelationships between the various dimensions. Indeed, an exclusive focus on any one dimension encourages a myopic analysis of software systems development and fails to take into account how, for example, the content of a software systems project is intimately connected with the context in which development occurs, the processes it involves and the people who perform them. Using the framework as an analytical guide facilitates a more holistic analysis across multiple levels of analysis, and avoids the narrow focus on individual dimensions and factors that has proved to be inadequate in the past.

The classificatory framework also has potential practical application in risk management. Consideration of the content, process and context of software systems projects, as well as their potential interaction, could form the basis of an inclusive project risk framework. For example, during initial risk identification, the various factors could provide a comprehensive list of areas of potential risk that need to be addressed. Further, the four dimensions of the framework could represent general themes around which these risks could be grouped for subsequent ongoing risk management as a software systems project proceeds. Inclusion of contextual elements, such as the organizational history of software systems development and use, facilitates the organizational learning from past projects that is a critical part of risk management.

The various themes outlined in the discussion section of this paper suggest that future work in this field needs to take several trends into account. Given the substantial changes observed over the past decade, it seems reasonable to expect further changes in software systems development and acquisition in the future. Empirical research needs to address these changes while attending to any persistent aspects. The continued occurrence of perceived systems project failures suggests that simple prescriptive lists of 'best practice' factors are inadequate, and that research is needed that acknowledges and explores the complex interrelationships and interactions between factors in software systems projects. The importance of people and process to project outcomes is likely to grow as the nature of software systems development and acquisition continues to change, requiring more detailed analyses of processual, political and behavioral factors involving an increasing range and diversity of stakeholders. Finally, recognition that software systems projects take place in specific contextual settings implies a need for empirical research that focuses on the institutional dimensions of software systems development, across multiple contexts and levels of analysis.

## ACKNOWLEDGEMENT

The authors would like to thank the referees for their constructive comments and suggestions, which have helped to shape the final version of this paper.

## REFERENCES

Akkermans, H. & van Helden, K. (2002). Vicious and virtuous cycles in ERP implementation: a case study of interrelations between critical success factors. *European Journal of Information Systems, 11*(1), 35-46.

Aladwani, A.M. (2000). IS project characteristics and performance: a Kuwaiti illustration. *Journal of Global Information Management, 8*(2), 50-57.

Aladwani, A.M. (2002). An integrated performance model of information systems projects. *Journal of Management Information Systems, 19*(1), 185-210.

Al-Karaghouli, W., Alshawi, S. & Fitzgerald, G. (2005). Promoting requirement identification quality: enhancing the human interaction dimension. *The Journal of Enterprise Information Management, 18*(2), 256-267.

Alvarez, R. (2002). Confessions of an information worker: a critical analysis of information requirements discourse. *Information and Organization, 12*(2), 85-107.

Amoako-Gyampah, K. (1997). Exploring users' desires to be involved in computer systems development: an exploratory study. *Computers in Human Behavior, 13*(1), 65-81.

Amoako-Gyampah, K. & White, K.B. (1997). When is user involvement not user involvement? *Information Strategy: The Executive's Journal, 13*(4), 40-45.

Asaro, P.M. (2000). Transforming society by transforming technology: the science and politics of participatory design. *Accounting, Management and Information Technologies, 10*(4), 257-290.

Avgerou, C. (2001). The significance of context in information systems and organizational change. *Information Systems Journal, 11*, 43-63.

Avison, D.E. & Fitzgerald, G. (2003). Where now for development methodologies? *Communications of the ACM, 46*(1), 79-82.

Baddoo, N., Hall, T. & Jagielska, D. (2006). Software developer motivation in a high maturity company: a case study. *Software Process Improvement and Practice, 11*(3), 219-228.

Bahli, B. & Tullio, D. (2003). Web engineering: an assessment of empirical research. *Communications of the AIS, 12*, 203-222.

Barki, H. & Hartwick, J. (1994). Measuring user participation, user involvement and user attitude. *MIS Quarterly, 18*(1), 59-82.

Barki, H., Rivard, S. & Talbot, J. (2001). An integrative contingency model of software project risk management. *Journal of Management Information Systems, 17*(4), 37-69.

Barry, C. & Lang, M. (2003). A comparison of 'traditional' and multimedia information systems development practices. *Information and Software Technology, 45*, 217-227.

Baskerville, R. & Pries-Heje, J. (2004). Short cycle time systems development. *Information Systems Journal, 14*(3), 237-264.

Beynon-Davies, P., Tudhope, D. & Mackay, H. (1999). Information systems prototyping in practice. *Journal of Information Technology, 14*(1), 107-120.

Bradley, J.H. & Hebert, F.J. (1997). The effect of personality type on team performance. *Journal of Management Development, 16*(5), 337-353.

Briggs, R.O., De Vreede, G.-J., Nunamaker, J.F. & Sprague, R.H. (2003). Special issue: information systems success. *Journal of Management Information Systems, 19*(4), 5-8.

Britton, C., Jones, S., Myers, M. & Sharif, M. (1997). A survey of current practice in the development of multimedia systems. Information and Software Technology, 39(10), 695-705.

Bussen, W. & Myers, M.D. (1997). Executive information systems failure: a New Zealand case study. *Journal of Information Technology, 12*, 145-153.

Butler, T. (2003). An institutional perspective on developing and implementing intranet and internet-based information systems. *Information Systems Journal, 13*(3), 209-231.

Butler, T. & Fitzgerald, B. (1997). A case study of user participation in the Information Systems process. In K. Kumar & J.I. DeGross (Eds.), *Proceedings of the 18th International Conference on Information Systems (Atlanta, GA)* (pp. 411-426). Atlanta, GA: Association of Information Systems.

Butler, T. & Fitzgerald, B. (1999a). The institutionalisation of user participation for systems development in Telecom Eireann. In M. Khosrowpour (Ed.), *Success and Pitfalls of Information Technology Management* (pp. 68-86). Hershey, USA: Idea Group Publishing.

Butler, T. & Fitzgerald, B. (1999b). Unpacking the systems development process: an empirical application of the CSF concept in a research context. *Journal of Strategic Information Systems, 8*(4), 351-371.

Butler, T. & Fitzgerald, B. (2001). The relationship between user participation and the management of change surrounding the development of information systems: a European perspective. *Journal of End User Computing, 13*(1), 12-25.

Chae, B. & Poole, M.S. (2005). The surface of emergence: agency, institutions, and large-scale information systems. *European Journal of Information Systems, 14*(1), 19-36.

Chang, H.H. (2006). Technical and management perceptions of enterprise information system importance, implementation and benefits. *Information Systems Journal, 16*(3), 263-292.

Charette, R.N. (2005). Why software fails. *IEEE Spectrum, 42*(9), 42-49.

Chatzoglou, P.D. (1997). Use of methodologies: an empirical analysis of their impact on the economics of the development process. *European Journal of Information Systems, 6*(4), 256-270.

Christiaanse, E. & Huigen, J. (1997). Institutional dimensions in information technology implementation in complex network settings. *European Journal of Information Systems, 6*(2), 77-85.

Clegg, C.W., Axtell, C., Damodaran, L., Farbey, B., Hull, R., Lloyd-Jones, R., Nicholls, J., Sell, R. & Tomlinson, C. (1997). Information technology: a study of performance and the role of human and organizational factors. *Ergonomics, 40*(9), 851-871.

Coakes, J.M. & Coakes, E.W. (2000). Specifications in context: stakeholders, systems and modelling of conflict. *Requirements Engineering, 5*(3), 103-133.

Constantinides, P. & Barrett, M. (2006). Negotiating ICT development and use: the case of a telemedicine system in the healthcare region of Crete. *Information and Organization, 16*(1), 27-55.

Coombs, C.R., Doherty, N.F. & Loan-Clarke, J. (1999). Factors affecting the level of success of community information systems. *Journal of Management in Medicine, 13*(3), 142-153.

Coughlan, J., Lycett, M. & Macredi, R.D. (2003). Communication issues in requirements elicitation: a content analysis of stakeholder experiences. *Information and Software Technology, 45*(2), 525-537.

Crowston, K., Howison, J. & Annabi, H. (2006). Information systems success in free and open source software development: theory and measures. *Software Process Improvement and Practice, 11*(2), 123-148.

DeLone, W.H. & McLean, E.R. (2003). The DeLone and Mclean of information systems success: a ten-year update. *Journal of Managment Information Systems, 19*(4), 9-30.

Dhillon, G. (2004). Dimensions of power and IS implementation. *Information & Management, 41*(5), 635-644.

Doherty, N.F. & King, M. (1998a). The consideration of organizational issues during the systems development process: an empirical analysis. *Behaviour and Information Technology, 17*(1), 41-51.

Doherty, N.F. & King, M. (1998b). The importance of organisational issues in systems development. *Information Technology and People, 11*(2), 104-123.

Doherty, N.F. & King, M. (2001). An investigation of the factors affecting the successful treatment of organisational issues in systems development projects. *European Journal of Information Systems, 10*, 147-160.

Doherty, N.F., King, M. & Al-Mushayt, O. (2003). The impact of the inadequacies in the treatment of organizational issues on information systems development projects. *Information and Management, 41*, 49-62.

Doolin, B. (1999). Sociotechnical networks and information management in health care. *Accounting, Management and Information Technologies, 9*(2), 95-114.

Doolin, B. (2004). Power and resistance in the implementation of a medical management information system. *Information Systems Journal, 14*(4), 343-362.

Drummond, H. (1996). The politics of risk: trials and tribulations of the Taurus project. *Journal of Information Technology, 11*(2), 347-357.

Eason, K. (2001). Changing perspectives on the organizational consequences of information technology. *Behaviour and Information Technology, 20*(5), 323-328.

Enquist, H. & Makrygiannis, N. (1998). Understanding misunderstandings. In *Proceedings of the Thirty-First Hawaii International Conference on System Sciences (6-9 January)* (Vol. 6, pp. 83-92). Kohala Coast, HI.

Espinosa, J.A., DeLone, W.H. & Lee, G. (2006). Global boundaries, task processes and IS project success: a field study. *Information Technology and People*, 19(4), 345-370.

Fitzgerald, B. (1998a). An empirical investigation into the adoption of systems development methodologies. *Information & Management, 34(6)*, 317-328.

Fitzgerald, B. (1998b). An empirically-grounded framework for the information systems development process. In R. Hirschheim, M. Newman & J.I. DeGross (Eds.), *Proceedings of the International Conference on Information Systems (Helsinki, Finland)* (pp. 103-114). Atlanta, GA: Association of Information Systems.

Fitzgerald, B. (1998c). A tale of two roles: the use of systems development methodologies in practice. In N. Jayaratna, A.T. Wood-Harper & B. Fitzgerald (Eds.), *Educating Methodology Practitioners and Researchers*. London: Springer-Verlag.

Fitzgerald, B. (2000). System development methodologies: the problem of tenses. *Information Technology and People, 13*(3), 174-185.

Fitzgerald, B. & Fitzgerald, G. (1999). Categories and contexts of information systems development: making sense of the mess. In C. Ciborra (Ed.), *Proceedings of the 7th European Conference of Information Systems* (pp. 194-211). Copenhagen, Denmark.

Fitzgerald, B., Russo, N.L. & Stolterman, E. (2002). *Information systems development: methods in action*. London: McGraw-Hill.

Flynn, D.J. & Jazi, M.D. (1998). Constructing user requirements: a social process for a social context. *Information Systems Journal, 8*(1), 53-83.

Foster, S.T. & Franz, C.R. (1999). User involvement during information systems development: a comparison of analyst and user perceptions of system acceptance. *Journal of Engineering and Technology Management, 16*(3-4), 329-348.

Galliers, R.D. & Swan, J.A. (2000). There's more to information systems development than structured approaches: information requirements analysis as a socially mediated process. *Requirements Engineering, 5*(2), 74-82.

Gallivan, M.J. & Keil, M. (2003). The user-developer communication process: a critical case study. *Information Systems Journal, 13*(1), 37-68.

Gärtner, J. & Wagner, I. (1996). Mapping actors and agendas: political frameworks of systems design and participation. *Human-Computer Interaction, 11*(3), 187-214.

Gasson, S. (1999). A social action model of situated IS design. *The Data Base for Advances in Information Systems, 30*(2), 82-97.

Gasson, S. (2006). A genealogical study of boundary-spanning IS design. *European Journal of Information Systems, 15*(1), 26-41

Goldstein, H. (2005). Who killed the Virtual Case File? *IEEE Spectrum, 42*(9), 24-35.

Gowan, J.A. & Mathieu, R.G. (2005). The importance of management practices in IS project performance. *The Journal of Enterprise Information Management, 18*(2), 235-255.

Guinan, P.J., Cooprider, J.G. & Faraj, S. (1998). Enabling software team performance during requirements definition: a behavioral versus technical approach. *Information Systems Research, 9*(2), 101-125.

Hardgrave, B.C., Wilson, R.L. & Eastman, K. (1999). Toward a contingency model for selecting an information system prototyping strategy. *Journal of Management Information Systems, 16*(2), 113-136.

Hartwick, J. & Barki, H. (2001). Communication as a dimension of user participation. *IEEE Transactions on Professional Communication, 44*(1), 21-31.

Heiskanen, A., Newman, M. & Similä, J. (2000). The social dynamics of software development. *Accounting, Management and Information Technologies, 10*(1), 1-32.

Hornik, S., Chen, H.-G., Klein, G. & Jiang, J.J. (2003). Communication skills of IS providers: an expectation gap analysis from three stakeholder perspectives. *IEEE Transactions on Professional Communication, 46*(1), 17-34.

Howcroft, D. & Light, B. (2006). Reflections on issues of power in packaged software selection. *Information Systems Journal, 16*(3), 215-235.

Howcroft, D. & Wilson, M. (2003). Participation: 'bounded freedom' or hidden constraints on user involvement. *New Technology, Work and Employment, 18*(1), 2-19.

Hunton, J.E. & Beeler, J.D. (1997). Effects of user participation in systems development: a longitudinal field experiment. *MIS Quarterly*, 359-388.

Hwang, M.I. & Thorn, R.G. (1999). The effect of user engagement on system success: a meta-analytical integration of research findings. *Information & Management, 35*(4), 229-236.

Iivari, J., Hirschheim, R. & Klein, H.K. (2000/2001). A dynamic framework for classifying information systems development methodologies and approaches.

*Journal of Management Information Systems, 17*(3), 179-218.

Iivari, J. & Igbaria, M. (1997). Determinants of user participation: a Finnish survey. *Behaviour and Information Technology, 16*(2), 11-121.

Iivari, J. & Maansaari, J. (1998). The usage of systems development methods: are we stuck to old practices? *Information and Software Technology, 40*(9), 501-510.

Iivari, N. (2004a). Enculturation of user involvement in software development organizations - an interpretive case study in the product development context. In *Proceedings of the Third Nordic Conference on Human-Computer Interaction (Tampere, Finland)* (pp. 287-296). ACM Press: New York.

Iivari, N. (2004b). Exploring the rhetoric on representing the user: discourses on user involvement in software development. In R. Agarwal, L.J. Kirsch & J.I. DeGross (Eds.), *Proceedings of the 25th International Conference on Information Systems (Washington DC, Dec 12-15)* (pp. 631-643).

Irani, Z., Sharif, A.M. & Love, P.E.D. (2001). Transforming failure into success through organisational learning: an analysis of a manufacturing information system. *European Journal of Information Systems, 10*, 55-66.

Jiang, J.J., Chen, E. & Klein, G. (2002). The importance of building a foundation for user involvement in information systems projects. *Project Management Journal, 33*(1), 20-26.

Jiang, J.J. & Klein, G. (1999). Risks to different aspects of system success. *Information & Management, 36(5)*, 263-272.

Jiang, J.J. & Klein, G. (2000). Software development risks to project effectiveness. *The Journal of Systems and Software, 52*(1), 3-10.

Jiang, J.J., Klein, G. & Balloun, J.L. (1996). Ranking of system implementation success factors. *Project Management Journal, 27*, 50-55.

Jiang, J.J., Klein, G. & Balloun, J.L. (1998a). Perceptions of software development failures. *Information and Software Technology, 39*(14-15), 933-937.

Jiang, J.J., Klein, G. & Balloun, J.L. (1998b). Systems analysts' attitudes towards information systems development. *Information Resources Management Journal, 11*(4), 5-10.

Jiang, J.J., Klein, G. & Chen, H.-G. (2006). The effects of user partnering and user non-support on project performance. *Journal of the Association for Information Systems, 7*(2), 68-90.

Jiang, J.J., Klein, G. & Discenza, R. (2002). Pre-project partnering impact on an information system project, project team and project manager. *European Journal of Information Systems, 11*(2), 86-97.

Jiang, J.J., Klein, G. & Means, T.L. (2000). Project risk impact on software development team performance. *Project Management Journal, 31*(4), 19-26.

Jiang, J.J., Sobol, M.G. & Klein, G. (2000). Performance ratings and importance of performance measures for IS staff: the different perceptions of IS users and IS staff. *IEEE Transactions on Engineering Management, 47*(4), 424-434.

Johnson, J., Boucher, K.D., Connors, K. & Robinson, J. (2001). The criteria for success. *Software Magazine, 21*(1), S3-S11.

Jonasson, I. (2002). Trends in developing web-based multimedia information systems. In M. Kirikova, J. Grundspenkis, W. Wojtkowski, W.G. Wojtkowski, S. Wrycza & J. Zupancic (Eds.), *Information Systems Development: Advances in Methodologies, Components and Management* (pp. 79-86). New York: Kluwer Academic.

Jones, M.C. & Harrison, A.W. (1996). IS project performance: an empirical appraisal. *Information & Management, 31*(2), 51-65.

Jurison, J. (1999). Software project management: the manager's view. *Communications of the AIS, 2* (Article 17).

Kappelman, L.A., McKeeman, R. & Zhang, L. (2006). Early warning signs of IT project failure: the dominant dozen *Information Systems Management, 23*(4), 31 - 36.

Karlsen, J.T., Andersen, J., Birkel, L.S. & Odegard, E. (2005). What characterizes successful IT projects. *International Journal of Information Technology & Decision Making, 4*(4), 525-540.

Kautz, K. (2004). The enactment of methodology: the case of developing a multimedia information system. In R. Agarwal, L.J. Kirsch & J.I. DeGross (Eds.), *Proceedings of the 25th International Conference on Information Systems (Washington DC, USA)* (pp. 671-684). Atlanta, GA: Association for Information Systems.

Kautz, K., Hansen, B. & Jacobsen, D. (2004). The utilization of information systems development methodologies in practice. *Journal of Information Technology Cases and Applications, 6*(4), 1-20.

Kautz, K. & Nielsen, P.A. (2004). Understanding the implementation of software process improvement innovations in software organizations. *Information Systems Journal, 14*(1), 3–22.

Keil, M., Cule, P., Lyytinen, K. & Schmidt, R. (1998). A framework for identifying software projects risks. *Communications of the ACM, 14*(11), 76-83.

Keil, M. & Robey, D. (2001). Blowing the whistle on troubled software projects. *Communications of the ACM, 44*(4), 87-93.

Keil, M. & Tiwana, A. (2006). Relative importance of evaluation criteria for enterprise systems: a conjoint study. *Information Systems Journal, 16*(3), 237-262.

Keil, M., Tiwana, A. & Bush, A. (2002). Reconciling user and project manager perceptions on IT project risk: a Delphi study. *Information Systems Journal, 12*(2), 103-119.

Kiely, G. & Fitzgerald, B. (2002). An investigation of the information systems development environment: the nature of development life cycles and the use of methods. In *Proceedings of the Eighth Americas Conference of Information Systems (Dallas)* (pp. 1289-1296): AIS.

Kiely, G. & Fitzgerald, B. (2003). An investigation of the use of methods within information systems development projects. In M. Korpela, R. Montealegre & A. Poulymenakou (Eds.), *Proceedings of the IFIP WG8.2 & WG9.4 Working Conference on Information Systems Perspectives and Challenges in the Context of Globalization (Athens), In Progress Research Papers* (pp. 187-198): IFIP.

Kim, H.-W. & Pan, S.L. (2006). Towards a process model of information systems implementation: the case of Customer Relationship Management (CRM). *The Data Base for Advances in Information Systems, 37*(1), 59-76.

Kim, C.S. & Peterson, D.K. (2003). A comparison of the perceived importance of information systems development strategies by developers from the United States and Korea. *Information Resources Management Journal, 16*(1), 1-18.

Kim, C.S., Peterson, D.K. & Kim, J.H. (1999/2000). Information systems success: perceptions of developers in Korea. *The Journal of Computer Information Systems, 40*(2), 90-95.

Kirsch, L.J. & Beath, C.M. (1996). The enactments and consequences of token, shared, and compliant participation in information systems development. *Accounting, Management and Information Technologies, 6*(4), 221-254.

Knights, D. & Murray, F. (1994). *Managers Divided: Organisation Politics and Information Technology Management*. Chichester: Wiley.

KPMG. (2005). *Global IT Project Management Survey*. Switzerland: KPMG International.

Krishna, S. & Walsham, G. (2005). Implementing public information systems in developing countries: learning from a success story. *Information Technology for Development, 11*(2), 123-140.

Kujala, S. (2003). User involvement: a review of the benefits and challenges. *Behaviour and Information Technology, 22*(1), 1-16.

Kumar, K., van Dissel, H.G. & Bielli, P. (1998). The Merchant of Prato - *revisited*: toward a third rationality of information systems. *MIS Quarterly, 22*(2), 199-226.

Lang, M. & Fitzgerald, B. (2005). Hypermedia systems development practices: a survey. *IEEE Software, 22*(2), 68-75.

Lang, M. & Fitzgerald, B. (2006). New branches, old roots: a study of methods and techniques in Web/hypermedia systems design. *Information Systems Management, 23*(3), 62-74.

Larman, C. & Basili, V.R. (2003). Iterative and incremental development: a brief history. *Computer, 36*(6), 47-56.

Lemon, W.F., Liebowitz, J., Burn, J.M. & Hackney, R. (2002). Information systems project failure: a comparative study of two countries. *Journal of Global Information Management, 10*(2), 28-39.

Li, E.Y. (1997). Perceived importance of information system success factors: a meta analysis of group differences. *Information & Management, 32*(1), 15-28.

Liebowitz, J. (1999). Information systems: success or failure? *Journal of Computer Information Systems, 40*(1), 17-26.

Lin, W.T. & Shao, B.B.M. (2000). The relationship between user participation and system success: a contingency approach. *Information & Management, 37*(6), 283-295.

Linberg, K.R. (1999). Software developer perceptions about software project failure. *The Journal of Systems and Software, 49*, 177-192.

Lu, H.-P. & Wang, J.-Y. (1997). The relationships between management styles, user participation, and system success over MIS growth stages. *Information & Management, 32*(4), 203-213.

Lucas, H.C. (1975). *Why Information Systems Fail*. New York: Columbia University Press.

Luna-Reyes, L.F., Zhang, J., Gil-Garcia, J.R. & Cresswell, A.M. (2005). Information systems development as emergent socio-technical change: a practice approach. *European Journal of Information Systems, 14*(1), 93-105.

Lynch, T. & Gregor, S. (2004). User participation in decision support systems development: influencing system outcomes. *European Journal of Information Systems, 13*, 286-301.

Lyytinen, K. & Hirschheim, R. (1987). Information systems failures: a survey and classification of the empirical literature. *Oxford Surveys in Information Technology, 4*, 257-309.

Lyytinen, K. & Robey, D. (1999). Learning failure in information systems development. *Information Systems Journal, 9*, 85-101.

Mabert, V.A., Soni, A. & Venkataramanan, M.A. (2003). Enterprise resource planning: managing the implementation process. European Journal of Operational Research, 146(2), 302-314

Mahaney, R.C. & Lederer, A.L. (2003). Information systems project management: an agency theory interpretation. *Journal of Systems and Software, 68*(1), 1-9.

Mahmood, M.A., Burn, J.M., Gemoets, L.A. & Jacquez, C. (2000). Variables affecting information technology end-user satisfaction: a meta-analysis of the empirical literature. *International Journal of Human-Computer Studies, 52*, 751-771.

Marion, L. & Marion, D. (1998). Information technology professionals as collaborative change agents: a case study of behavioral health care. *Bulletin of the American Society for Information Science, 24*(6), 9-12.

Markus, M.L. & Benjamin, R.I. (1996). Change agentry - the next IS frontier. *MIS Quarterly, 20*(4), 385-407.

Markus, M.L. & Mao, J.-Y. (2004). Participation in development and implementation - updating an old, tired concept for today's IS contexts. *Journal of the Association for Information Systems, 5*(11-12), 514-544.

Markus, M.L. & Robey, D. (1988). Information technology and organizational change: causal structure in theory and research. *Management Science, 34*(5), 583-598.

Martin, A. & Chan, M. (1996). Information systems project redefinition in New Zealand: will we ever learn? *The Australian Computer Journal, 28*(1), 27-40.

McKeen, J.D. & Guimaraes, T. (1997). Successful strategies for user participation in systems development. *Journal of Management Information Systems, 14*(2), 133-150.

Mitev, N. (2000). Towards social constructivist understandings of ISD success and failure: introducing a new computerised reservation system. In W.J. Orlikowski, S. Ang, P. Weill, H.C. Krcmar & J.I. DeGross (Eds.), *Proceedings of the Twenty-First International Conference on Information Systems* (pp. 84-93): Association for Information Systems.

Myers, M.D. & Young, L.W. (1997). Hidden agendas, power and managerial assumptions in information systems development: an ethnographic study. *Information Technology and People, 10*(3), 224-240.

Nandhakumar, J. (1996). Design for success?: critical success factors in executive information systems development. *European Journal of Information Systems, 5*(1), 62-72.

Nandhakumar, J. & Avison, D.E. (1999). The fiction of methodical development: a field study of information systems development. *Information Technology and People, 12*(2), 176-191.

Nandhakumar, J. & Jones, M. (1997). Designing in the dark: the changing user-developer relationship in information systems development. In K. Kumar & J.I. DeGross (Eds.), *Proceedings of the Eighteenth International Conference on Information Systems (Atlanta, Georgia, USA)* (pp. 75-88). Atlanta, GA: Association for Information Systems.

Nelson, R.R. (2005). Project retrospectives: evaluating project success, failure and everything in between. *MIS Quarterly Executive, 4*(3), 361-372.

Newman, M. & Sabherwal, R. (1996). Determinants of commitment to information systems development: a longitudinal investigation. *MIS Quarterly, 20*, 23-54.

Nicolaou, A.I. (1999). Social control in information systems development. *Information Technology and People, 12*(2), 130-147.

Olesen, K. & Myers, M.D. (1999). Trying to improve communication and collaboration with information technology: an action research project which failed. *Information Technology and People, 12*(4), 317-332.

Oz, E. & Sosik, J.J. (2000). Why information systems projects are abandoned: a leadership and communication theory and exploratory study. *Journal of Computer Information Systems, 44*(1), 66-78.

Pan, G.S.C. (2005). Information systems project abandonment: a stakeholder analysis. *International Journal of Information Management, 25*, 173-184.

Pan, G.S.C. & Flynn, D.J. (2003). Information systems project abandonment: a case of political influence by the stakeholders. *Technology Analysis & Strategic Management, 15*(4), 457-466.

Pan, G.S.C., Pan, S.L. & Flynn, D.J. (2004). De-escalation of commitment to information systems projects: a process perspective. *Journal of Strategic Information Systems, 13*, 247-270.

Parr, A. & Shanks, G. (2000). A model of ERP project implementation. *Journal of Information Technology, 15*(4), 289–303.

Peterson, D.K. & Kim, C.S. (2003). Perceptions on IS risks and failure types: a comparison of designers from the United States, Japan and Korea. *Journal of Global Information Management, 11*(2), 19-38.

Peterson, D.K., Kim, C.S., Kim, J.H. & Tamura, T. (2002). The perceptions of information systems designers from the United States, Japan, and Korea on success and failure factors. *International Journal of Information Management, 22*(6), 421-439.

Pouloudi, A. & Whitley, E.A. (1997). Stakeholder identification in inter-organizational systems: gaining insights for drug use management systems. *European Journal of Information Systems, 6*(1), 1-14.

Poulymenakou, A. & Holmes, A. (1996). A contingency framework for the investigation of information systems failure. *European Journal of Information Systems, 5*(1), 34-46.

Procaccino, J.D. & Verner, J.M. (2006). Software project managers and project success: an exploratory study. *The Journal of Systems and Software, 79*(11), 1541–1551.

Procaccino, J.D., Verner, J.M., Darter, M.E. & Amadio, W.J. (2005). Toward predicting software development success from the perspective of practitioners: an exploratory Bayesian model. *Journal of Information Technology, 20*(3), 187-200

Procaccino, J.D., Verner, J.M. & Lorenzet, S.J. (2006). Defining and contributing to software development success. *Communications of the ACM, 49*(8), 79-83.

Ravichandran, T. & Rai, A. (2000). Quality management in systems development: an organizational system perspective. *MIS Quarterly, 24*(3), 381-415.

Reel, J.S. (1999). Critical success factors in software projects. *IEEE Software, 16*(3), 18-23.

Riley, L. & Smith, G. (1997). Developing and implementing IS: a case study analysis in social services. *Journal of Information Technology, 12*(4), 305-321.

Roberts, T.L., Leigh, W. & Purvis, R.L. (2000). Perceptions on stakeholder involvement in the implementation of system development methodologies. *Journal of Computer Information Systems, 40*(3), 78-83.

Robey, D. & Boudreau, M.-C. (1999). Accounting for the contradictory organizational consequences of information technology: theoretical directions and methodological implications. *Information Systems Research, 10*(2), 167-185.

Robey, D. & Newman, M. (1996). Sequential patterns in information systems development: an application of a social process model. *ACM Transaction on Information Systems, 14*(1), 30-63.

Robey, D., Welke, R.J. & Turk, D. (2001). Traditional, iterative, and component-based development: asocial analysis of software development paradigms. *Information Technology and Management, 2*(1), 53-70.

Royal Academy of Engineering. (2004). *The Challenges of Complex IT Projects*. London: Royal Academy of Engineering.

Saleem, N. (1996). An empirical test of the contingency approach to user participation in information systems development. *Journal of Management Information Systems, 13*(1), 145-166.

Sarkkinen, J. & Karsten, H. (2005). Verbal and visual representations in task redesign: how different viewpoints enter into information systems design discussions. *Information Systems Journal, 15*(3), 181-211.

Sauer, C. (1999). Deciding the future for IS failures: not the choice you might think. In R.D. Galliers & W.L. Currie (Eds.), *Rethinking Management Information Systems: An Interdisciplinary Perspective* (pp. 279-309). Oxford: Oxford University Press.

Sawyer, S. (2001a). Effects of intra-group conflict on packaged software development team performance. *Information Systems Journal, 11*, 155-178.

Sawyer, S. (2001b). A market-based perspective on information systems development. *Communications of the ACM, 44*(11), 97-102.

Sawyer, S. & Guinan, P.J. (1998). Software development: processes and performance. *IBM Systems Journal, 37*(4), 552–569.

Schmidt, R., Lyytinen, K., Keil, M. & Cule, P. (2001). Identifying software project risks: an international Delphi study. *Journal of Management Information Systems, 17*(4), 5-36.

Scott, J.E. & Vessey, I. (2002). Managing risks in enterprise systems implementation. *Communications of the ACM, 45*(4), 74-81.

Serafeimidis, V. & Smithson, S. (1999). Rethinking the approaches to information systems investment evaluation. *Logistics Information Management, 12*(1/2), 94–107.

Sharma, R. & Yetton, P. (2003). The contingent effects of management support and task interdependence on successful information systems implementation. *MIS Quarterly, 27*(4), 533-555.

Skok, W. & Legge, M. (2002). Evaluating enterprise resource planning (ERP) systems using an interpretive approach. *Knowledge and Process Management, 9*(2), 72-82.

Software Magazine. (2004). *Standish: Project Success Rates Improved Over 10 Years*. Retrieved 6 August, 2004, from http://www.softwaremag.com/L.cfm?Doc=newsletter/2004-01-15/Standish

Somers, T.M. & Nelson, K. (2001). The impact of critical success factors across stages of Enterprise Resource Planning implementations. In *Proceedings of the 34th Hawaii International Conference on System Sciences* (Vol. 8, pp. 8016). Washington, DC: IEEE Computer Society.

Somers, T.M. & Nelson, K. (2004). A taxonomy of players and activities across the ERP project life cycle. Information & Management, 41(3), 257-278.

Standing, C., Guilfoyle, A., Lin, C. & Love, P.E.D. (2006). The attribution of success and failure in IT projects. Industrial Management & Data Systems, 100(8), 1148-1165.

Standish Group International. (1999). *CHAOS: A Recipe for Success (1998)*. West Yarmouth, Massachusetts: The Standish Group International, Inc.

Standish Group International. (2001). *Extreme CHAOS (2000)*. West Yarmouth, Massachusetts: The Standish Group International, Inc.

Staples, D.S., Wong, I. & Seddon, P.B. (2002). Having expectations of information systems benefits that match received benefits: does it really matter? *Information & Management, 40(2)*, 115-131.

Stockdale, R. & Standing, C. (2006). An interpretive approach to evaluating information systems: a content, context, process framework. *European Journal of Operational Research, 173*(3), 1090–1102.

Sumner, M. (2000). Risk factors in enterprise-wide/ERP projects. *Journal of Information Technology, 15*(4), 317-327.

Sumner, M., Bock, D. & Giamartino, G. (2006). Exploring the linkage between the characteristics of it project leaders and project success *Information Systems Management, 23*(4), 43-49.

Symon, G. (1998). The work of IT system developers in context: an organizational case study. *Human-Computer Interaction, 13*(1), 37-71.

Symon, G. & Clegg, C.W. (2005). Constructing identity and participation during technological change. *Human Relations, 58*(9), 1141-1161.

Taylor, M.J., McWilliam, J., Forsyth, H. & Wade, S. (2002). Methodologies and website development: a survey of practice. *Information and Software Technology, 22*, 381-391.

Taylor-Cummings, A. (1998). Bridging the user-IS gap: a study of major information systems projects. *Journal of Information Technology, 13*(1), 29-54.

Terry, J. & Standing, C. (2004). The value of user participation in e-commerce systems development. *Informing Science, 7*, 31-46.

Umble, E.J., Haft, R.R. & Umble, M.M. (2003). Enterprise resource planning: implementation procedures and critical success factors. *European Journal of Operational Research, 146*(2), 241-257.

Urquhart, C. (1999). Themes in early requirements gathering: the case of the analyst, the client and the student assistance scheme. *Information Technology and People, 12*(1), 44-70.

Urquhart, C. (2001). Analysts and clients in organisational contexts: a conversational perspective. *Strategic Information Systems, 10*, 243-262.

van Offenbeek, M.A.G. & Koopman, P.L. (1996). Information systems development: from user participation to contingent interaction among involved parties. *European Journal of Work and Organizational Psychology, 5*(3), 421-438.

Verner, J.M. & Evanco, W.M. (2005). In-house software development: what project management practices lead to success? *IEEE Software, 22*(1), 86-93.

Vidgen, R. (2002). Constructing a web information system development methodology. *Information Systems Journal, 12*(3), 247-261.

Vidgen, R., Madsen, S. & Kautz, K. (2004). Mapping the information systems development process. In B. Fitzgerald & E.H. Wynn (Eds.), *IT Innovation for Adaptability and Competitiveness* (pp. 157-172). Boston: Kluwer Academic Press.

Vinekar, V., Slinkman, C.W. & Nerur, S. (2006). Can agile and traditional systems development approaches coexist? An ambidextrous view. *Information Systems Management, 23*(3), 31-42.

Wallace, L. & Keil, M. (2004). Software project risks and their effect on outcomes. *Communications of the ACM, 47*(4), 68-73.

Walsham, G. (1993). *Interpreting Information Systems in Organizations*. Chichester: John Wiley and Sons.

Walsham, G. (2002). Cross-cultural software production and use: a structurational analysis. *MIS Quarterly, 26*(4), 359-380.

Wang, E.T.G., Chou, H.-W. & Jiang, J.J. (2005). The impacts of charismatic leadership style on team cohesiveness and overall performance during ERP implementation. *International Journal of Project Management, 23*(2), 173–180.

Wang, E.T.G., Shih, S.-P., Jiang, J.J. & Klein, G. (2006). The relative influence of management control and user-IS personnel interaction on project performance. *Information and Software Technology, 48*(3), 214-220.

Warne, L. & Hart, D. (1996). The impact of organizational politics on information systems project failure-a case study. In *Proceedings of the Twenty-Ninth Hawaii International Conference on System Sciences* (Vol. 4, pp. 191-201): IEEE.

Wastell, D. & Newman, M. (1996). Information system design, stress and organisational change in the ambulance services: a tale of two cities. *Accounting, Management and Information Technologies, 6*(4), 283-300.

Wiersema, M.F. & Bantel, K.A. (1992). Top management team demography and corporate strategic change. *Academy of Management Journal, 35*(1), 91-121.

Williams, L. & Cockburn, A. (2003). Agile software development: it's about feedback and change. *Computer, 36*(6), 39-43.

Wilson, M. (2002). Making nursing visible? Gender, technology and the care plan script. *Information Technology and People, 15*(2), 139-158.

Wilson, M. & Howcroft, D. (2000). The politics of IS evaluation: a social shaping perspective. In W.J. Orlikowski, S. Ang, P. Weill, H.C. Krcmar & J.I. DeGross (Eds.), *Proceedings of the Twenty-First International Conference on Information Systems* (pp. 94-103): Association for Information Systems.

Wilson, M. & Howcroft, D. (2002). Re-conceptualising failure: social shaping meets IS research. *European Journal of Information Systems, 11*, 236-250.

Wilson, S., Bekker, M., Johnson, P. & Johnson, H. (1997). Helping and hindering user involvement - a tale of everyday design. In S. Pemberton (Ed.), *Proceedings of the SIGCHI Conference on Human Factors in Computing Systems (22-27 March, Atlanta, GA)* (pp. 178-185): ACM Press.

Wixom, B. & Watson, H.J. (2001). An empirical investigation of the factors affecting data warehousing success. *MIS Quarterly, 25*(1), 17-41.

Wynekoop, J.L. & Russo, N.L. (1997). Studying system development methodologies: an examination of research methods. *Information Systems Journal, 7*(1), 47-65.

Yetton, P., Martin, A., Sharma, R. & Johnston, K. (2000). A model of information systems project performance. *Information Systems Journal, 10*(4), 263-289.

Zeffane, R. & Cheek, B. (1998). Does user involvement during information systems development improve data quality? *Human Systems Management, 17*(2), 115-121.

**Appendix A:** Empirical studies reporting multiple factors influencing software system project outcomes. ('+' or '-' respectively indicate a positive or negative relationship to the project outcome measure. Italicized factors indicate a statistically significant relationship at or below a value of p=0.10.)

| Study | Focus | Method | People and action | Project content | Development processes | Institutional context | Measure of success (failure) |
|---|---|---|---|---|---|---|---|
| Aladwani [2000] | Systems project performance | Survey Kuwait | Project staff expertise (+) <br> *Top management support* (+) <br> Project team conflict (-) <br> Horizontal coordination (+) | Project complexity (-) <br> *Adequate development tools* (+) | *Project planning* (+) <br> *User participation* (+) | | Project efficiency <br> Project effectiveness |
| Barry & Lang [2003] | Multimedia systems project performance | Survey Ireland | Inadequate staff skills (-) <br> Unrealistic user expectations (-) | Scope creep (-) <br> Project complexity (-) <br> Staff shortages (-) <br> Cost overruns (-) <br> Time overruns (-) | Unclear requirements (-) <br> Lack of standard method (-) | | Not defined |
| Bussen & Myers [1997] | EIS implementation | Case study NZ | Lack of user commitment (-) <br> Lack of user readiness (-) <br> Lack of top management support (-) <br> Lack of user-developer communication (-) <br> Organizational politics (-) | Non-alignment with business goals (-) <br> Time overruns (-) <br> Staff turnover (-) <br> Technical problems (-) <br> Data problems (-) | Poorly defined requirements (-) <br> Lack of project planning (-) | Hierarchical organizational structure (-) <br> Changes in company ownership (-) <br> Rapid organizational growth (-) <br> Economic context (-) | Project abandonment |
| Butler & Fitzgerald [1999b] | Systems development critical success factors | Case studies Ireland | Top management support (+) <br> Adequate vendor support (+) | Use of prototyping tools (+) <br> Technical problems (-) | Well-defined requirements (+) <br> Project management (+) <br> Project planning (+) <br> User participation (+) <br> Use of a standard method (+) <br> Management of change (+) | | Not defined |
| Clegg et al. [1997] | Systems project performance | Interviews UK | Lack of project staff expertise (-) <br> Unrealistic user expectations (-) <br> Lack of top management support (-) <br> Organizational politics (-) | Non-alignment with business goals (-) <br> Cost overruns (-) <br> Time overruns (-) <br> Project complexity (-) | Unclear requirements (-) <br> Poor project management (-) <br> Inadequate standard methods (-) <br> Lack of user participation (-) <br> Adequate user training (+) <br> Poor management of change (-) | | Meeting system objectives |
| Irani et al. [2001] | MRP system implementation | Case study UK | User resistance (-) <br> Lack of top management support (-) <br> Use of external consultants (+) <br> Issues with external vendors (-) | Alignment with business goals (+) <br> Technological issues (-) | Project management (+) <br> Project management techniques (+) <br> Management of change (+) <br> User participation (+) <br> User training (+) | Organizational culture (+) | System use |
| Jiang et al. [1996] | Systems implementation | Survey US | User commitment (+) <br> Project staff expertise (+) <br> Top management support (+) <br> User-developer communication (+) | Clear project goals (+) <br> Adequate resources (+) <br> Appropriate technology (+) | Project management (+) <br> Project leadership (+) <br> User participation (+) | | Not defined |

| Study | Focus | Method | People and action | Project content | Development processes | Institutional context | Measure of success (failure) |
|---|---|---|---|---|---|---|---|
| Jiang et al. [1998a] | Systems development problems | Survey US | Lack of project staff communication skills (-)<br>Lack of project staff expertise (-) | Restricted scope (-)<br>Unclear project goals (-)<br>*Inadequate resources* (-)<br>Cost overruns (-)<br>Time overruns (-) | *Unclear requirements* (-)<br>Lack of project planning (-)<br>*Inadequate documentation* (-)<br>*Inadequate testing* (-)<br>Lack of user participation (-)<br>Lack of management of change (-)<br>*Lack of user training* (-) | | Not defined |
| Jiang & Klein [1999, 2000] | Systems project risks | Survey US | *Lack of user commitment* (-)<br>*Lack of user experience* (-)<br>Lack of project staff expertise (-)<br>Lack of project staff domain knowledge (-)<br>*Lack of functioning of project team* (-)<br>*Unclear role definition* (-)<br>Project team conflict (-) | Project size (-)<br>*Project complexity* (-)<br>*Technological newness* (-)<br>Inadequate resources (-) | Extent of change (-) | | Project effectiveness<br>Satisfaction with system<br>Organizational impact |
| Kappelman et al. [2006] | Systems project risks | Survey US | Lack of project staff expertise (-)<br>Lack of top management support (-)<br>Lack of project team commitment (-)<br>Poor communication (-) | Unclear business case (-)<br>Lack of resources (-)<br>Unavailability of appropriate expertise (-) | Lack of documented requirements (-)<br>Poor project planning (-)<br>Poor project leadership (-)<br>Lack of user participation (-)<br>Lack of management of change (-) | | Not defined |
| Keil et al. [2002] | Software project risks (users) | Delphi study US | User resistance (-)<br>Lack of project staff expertise (-)<br>Unclear role definition (-)<br>Poor project team relationships (-)<br>Organizational conflict (-) | Changing scope (-)<br>Inadequate/inappropriate project staff (-) | Misunderstanding requirements (-)<br>Lack of project planning (-)<br>Lack of user participation (-)<br>Non-use of a standard method (-) | | Not defined |
| Keil et al. [1998]; Schmidt et al. [2001] | Software project risks (project managers) | Delphi study Hong Kong US Finland | Unmanaged user expectations (-)<br>Lack of user commitment (-)<br>Lack of project staff expertise (-)<br>Lack of top management support (-)<br>Unclear role definition (-)<br>Organizational conflict (-) | Changing scope (-)<br>Inadequate/inappropriate project staff (-)<br>Staff turnover (-)<br>Technological newness (-) | Misunderstanding requirements (-)<br>Changing requirements (-)<br>Lack of user participation (-)<br>Lack of project leadership (-)<br>Poor management of change (-) | | Not defined |
| Kim & Pan [2006] | CRM systems critical success factors | Case studies Singapore | Top management support (+)<br>Project champion continuity (+)<br>Project team skills (+) | Adequate resources (+) | Requirements management (+)<br>Effective system design & development (+)<br>User participation (+)<br>Management of change (+)<br>Business process design (+) | | IS quality<br>User satisfaction<br>Use<br>Net benefits |
| Kumar et al. [1998] | Inter-organizational | Case study Italy | User commitment (+)<br>Support of institutional stakeholders (+) | Alignment with business goals (+) | Clear requirements (+)<br>Adequate user training (+) | Inattention to national cultural context (-) | Not defined |

| Study | Focus | Method | People and action | Project content | Development processes | Institutional context | Measure of success (failure) |
|---|---|---|---|---|---|---|---|
| | system implementation | | | Adequate resources (+)<br>Proven technology (+) | Adequate testing (+) | | |
| Lemon et al. [2002] | Systems project performance | Survey Australia US | Realistic user expectations (+)<br>Unmanaged user expectations (-)<br>Project staff expertise (+)<br>Top management support (+) | Clear project goals (+)<br>Inadequate/inappropriate project staff (-)<br>Inappropriate technology (-) | Clear requirements (+)<br>Lack of project management (-)<br>Project planning (+)<br>Small project milestones (+)<br>Lack of risk management (-)<br>Lack of project governance (-)<br>User participation (+) | Corporate culture (-) | Not defined |
| Linberg [1999] | Software project performance | Case study US | Lack of top management support (-)<br>Functioning of the project team (+)<br>Organizational conflict & politics (-) | Under-estimated project size (-)<br>Project complexity (-)<br>Low project importance (-)<br>Inadequate resources (-)<br>Time overruns (-)<br>Cost overruns (-)<br>Inadequate development tools (-) | Project management (+)<br>Project leadership (+)<br>Use of a standard method (+) | | Project completion or project cancellation |
| Mabert et al. [2003] | ERP systems implementation | Survey US | *Top management support (+)*<br>*Communication with external stakeholders (+)*<br>*Communication with internal stakeholders (-)* | Project size & complexity (-)<br>Major software modification (-) | *Project planning (+)*<br>Benchmarked progress against milestones (+)<br>Empowered decision-makers (+)<br>*Accelerated implementation strategy (+)*<br>*User training (+)*<br>Minor alignment of business processes (+) | | Completed on time and to budget |
| Martin & Chan [1996] | Systems project performance | Survey NZ | User resistance (-)<br>Lack of project staff expertise (-)<br>*Lack of top management support (-)*<br>Project champion (+)<br>Project team conflict (-)<br>Organizational politics (-)<br>Poor communication (-) | Project size (-)<br>Project complexity (-)<br>*Project newness (-)*<br>Project importance (+)<br>*Clear project scope & goals (+)*<br>Alignment with business goals (+)<br>Adequate resources (+)<br>Inadequate/inappropriate staff (-)<br>Cost overruns (-)<br>Time overruns (-)<br>Staff turnover (-)<br>Inappropriate technology (-) | Changing requirements (-)<br>Inexperienced project leader (-)<br>*Realistic project plan & schedule (+)*<br>*Allowance for developer learning (+)*<br>*Cost-benefit analysis (+)*<br>Use of standard methods (+)<br>Lack of user participation (-)<br>Lack of user training (-)<br>Lack of management of change (-) | *Growing IT function (-)*<br>*Focus on assimilating new technology (-)*<br>Organizational restructuring (-) | Project smoothly completed, redefined or abandoned |
| Nandhakumar [1996] | EIS critical success factors | Case study Europe | Top management support (+) | Alignment with business goals (+)<br>Availability of resources (+)<br>Use of appropriate technology (+) | Project planning (+) | Hierarchical organizational structure (-)<br>Organizational policy on resource allocation (+)<br>Poor market conditions (-) | Not defined |

| Study | Focus | Method | People and action | Project content | Development processes | Institutional context | Measure of success (failure) |
|---|---|---|---|---|---|---|---|
| | | | | Management of data problems (+) | | | |
| Oz & Sosik [2000] | Systems project performance | Survey US Europe | Lack of project staff expertise (-) Lack of top management support (-) Lack of functioning of the project team (-) Organizational politics (-) | Unclear project goals (-) Time overruns (-) Unrealistic deadlines (-) Cost overruns (-) Inappropriate technology (-) | Changing requirements (-) Poor project management (-) | | Project abandonment |
| Parr & Shanks [2000] | ERP systems critical success factors | Case studies Australia | Top management support (+) Project champion (+) Balanced project team mix (+) | Clear project scope and goals (+) Small project scope (+) Availability of appropriate project staff (+) Realistic deadlines (+) Minimal software modification (+) | Empowered decision-makers (+) Commitment to change (+) | | Completed on time and to budget |
| Procaccino et al. [2006] | Software project performance | Survey US | Realistic user expectations (+) Project staff expertise (+) Functioning of the project team (+) User-developer interaction (+) | Lack of project scope creep (+) Availability of appropriate expertise (+) | Clear requirements (+) Realistic & achievable requirements (+) Project management (+) Use of standard method (+) User participation (+) | | Not defined |
| Peterson et al. [2002]; Kim & Peterson [2003] | Systems project performance | Survey US Korea Japan | Project staff expertise (+) Top management support (+) | Appropriate project scope (+) Clear project goals (+) Alignment with business goals (+) Appropriate technology (+) | Project planning (+) Peer review & feedback (+) Project leadership (+) Use of standard method (+) User participation (+) Management of change (+) | | Not defined |
| Skok & Legge [2002] | ERP systems critical success factors | Case studies Europe | User resistance (-) Project staff expertise (+) Lack of developer domain knowledge (-) Top management support (+) Effective management of external consultants (+) User-developer communication (+) Conflict management (+) | Inadequate/inappropriate project staff (-) Staff turnover (-) | Experienced project leader (+) User participation (+) User training (+) Management of change (+) | Differences in national cultural context (-) | Not defined |
| Somers & Nelson [2001] | ERP systems critical success factors | Survey US | Management of user expectations (+) Project staff expertise (+) Top management support (+) Project champion (+) Adequate vendor support (+) Use of external consultants (+) | Clear project goals (+) Adequate resources (+) Use of appropriate software (+) High quality data sources (+) Compatible IT architecture (+) Minimal software modification (+) | Project management (+) Appropriate project governance (+) User training (+) Management of change (+) Alignment of business | Organizational culture of cooperation (+) | Not defined |

| Study | Focus | Method | People and action | Project content | Development processes | Institutional context | Measure of success (failure) |
|---|---|---|---|---|---|---|---|
| | | | Communication (+) | | processes (+) | | |
| Sumner [2000] | ERP systems project risks | Case studies US | Lack of developer domain knowledge (+) Lack of top management support (-) Lack of project champion (-) Ineffective use of external consultants (-) Ineffective communication (-) | Inadequate/inappropriate project staff (-) Staff turnover (-) Major software modification (-) Lack of data integration (-) Integration with legacy systems (-) Incompatible IT architecture (-) | Lack of centralized project management (-) Lack of user training (-) Lack of user participation (-) | | Not defined |
| Standish Group International [2001]; Johnson et al. [2001] | Systems project performance | Survey US | Top management support (+) | Clear business objectives (+) Minimized scope (+) Standard software infrastructure (+) | Stable requirements (+) User participation (+) Project leadership (+) Project planning (+) Use of a project management method (+) | | Completed on time, to budget and to specifications |
| Wastell & Newman [1996] | Systems project performance | Case studies UK | Top management support (+) Consultative management style (+) | Adequate resources (+) Proven technology (+) | Human-centered design approach (+) User participation (+) User training (+) Adequate testing (+) | Good industrial relations (+) | Service level improvement Project abandonment |
| Wixom & Watson [2001] | Data warehouse systems implementation | Survey US | *Project staff skills (+) Top management support (+) Project champion (+)* | *Adequate resources (+) Appropriate technology (+) High quality data sources (+)* | *User participation (+) Management of change (+)* | | Implementation success System success |
| Yetton et al. [2000] | Systems project performance | Survey UK NZ | *Top management support (+) Project team conflict (-)* | *Project size (-) Project newness (-) Project importance (+) Project technical risk (-) Staff turnover (-)* | *Project planning (+) User participation (+)* | | Project completion Budget variances |